\documentclass[11pt]{article}

\usepackage{epsfig}

\long\def\comment #1\commentend{} \long\def\commabs #1\commabsend{}
\long\def\commful #1\commfulend{#1}

\newtheorem{theorem}{Theorem}[section]

\newtheorem{claim}[theorem]{Claim}
\newtheorem{lemma}[theorem]{Lemma}

\newtheorem{corollary}[theorem]{Corollary}

\newtheorem{example}[theorem]{Example}

\def\inline#1:{\par\vskip 7pt\noindent{\bf #1:}\hskip 10pt}
\def\midinline#1:{\par\noindent{\bf #1:}\hskip 10pt}
\def\dnsinline#1:{\par\vskip -7pt\noindent{\bf #1:}\hskip 10pt}
\def\ddnsinline#1:{\newline{\bf #1:}\hskip 10pt}
\def\largeinline#1:{\par\vskip 7pt\noindent{\large\bf #1:}\hskip 10pt}

\def\cD{{\cal D}}

\def\cS{{\cal S}}
\def\cM{{\cal M}}

\def\blackslug{\hbox{\hskip 1pt \vrule width 4pt height 8pt
    depth 1.5pt \hskip 1pt}}
\def\QED{\quad\blackslug\lower 8.5pt\null\par}

\newcommand{\func}{F}

\newcommand{\MC}{{\cal MC}}

\newcommand{\TS}{{\cal ST}}
\newcommand{\LS}{{\cal LS}}

\def\Reset{\mbox{\sc Reset}}

\def\WW{\mbox{\sc WeightWatch}}

\def\CW{\mbox{\sc ChangeWatch}}

\def\FS{\mbox{\sc FSDL}}
\def\iFS{\mbox{\sc FSDL$^{k_i}_{p_i}$}}

\def\S{\mbox{\sc SDL}}

\def\DL{\mbox{\sc DL}}
\def\sS{\mbox{\sc Sem-DL}}

\def\SDFS{\mbox{\sc StatDFS}}

\def\ninit{n_{0}}

\def\vecn{{\bar{n}}}
\begin{document}
%%%%%%%%%%%%%%%%%%%%%%%

\title{General Compact Labeling Schemes for Dynamic Trees\\}
\author{
Amos Korman
\thanks{Information Systems Group, Faculty of IE\&M, The Technion,
Haifa, 32000 Israel.  E-mail: {\tt pandit@tx.technion.ac.il}.
Supported in part at the Technion by an Aly Kaufman fellowship.}
 }

\date{\today}

%\begin{document}
\begin{titlepage}
\def\thepage{}
\maketitle

\begin{abstract}
Let $F$ be a function on pairs of vertices. An {\em $F$- labeling
scheme} is composed of a {\em marker} algorithm for labeling the
vertices of a graph with short labels, coupled with a {\em decoder}
algorithm allowing one to compute $F(u,v)$ of any two vertices $u$
and $v$ directly from their labels. As applications for labeling
schemes concern mainly large and dynamically changing networks, it
is of interest to study {\em distributed dynamic} labeling schemes.
This paper investigates labeling schemes for dynamic trees. We
consider two dynamic tree models, namely, the {\em leaf-dynamic}
tree model in which at each step a leaf can be added to or removed
from the tree and the {\em leaf-increasing} tree model in which the
only topological event that may occur is that a leaf joins the tree.

A general method for constructing labeling schemes for dynamic trees
(under the above mentioned dynamic tree models) was previously
developed in  \cite{KPR04}. This method is based on extending an
existing {\em static} tree labeling scheme to the dynamic setting.
This approach fits many natural functions on trees, such as
distance, separation level, ancestry relation, routing (in both the
adversary and the designer port models), nearest common ancestor
etc.. Their resulting dynamic schemes incur
 overheads (over the static scheme)
on the label size and on the communication complexity. In
particular, all their schemes yield a multiplicative overhead factor
of $\Omega(\log n)$ on the label sizes of the static schemes.
Following \cite{KPR04}, we develop a different general method for
extending static labeling schemes to the dynamic tree settings. Our
method fits the same class of tree functions. In contrast to the
above paper, our trade-off is designed to minimize the label size,
sometimes at the expense of communication.

Informally, for any function $k(n)$ and any static $F$-labeling
scheme on trees, we present an $F$-labeling scheme on dynamic trees
incurring multiplicative overhead factors (over the static scheme)
of $O(\log_{k(n)} n)$ on the label size and $O(k(n)\log_{k(n)} n)$
on the amortized message complexity. In particular,  by setting
$k(n)=n^{\epsilon}$ for any $0<\epsilon<1$, we obtain dynamic
labeling schemes with asymptotically optimal label sizes and
sublinear amortized message complexity for the ancestry relation,
the id-based and label-based nearest common ancestor relation and
the routing function.
\end{abstract}

\end{titlepage}
\pagenumbering{arabic}

\section{Introduction}
\paragraph*{\bf Motivation:}
Network representations have played an extensive and often crucial
role in many domains of computer science, ranging from data
structures, graph algorithms to distributed computing and
communication networks. Research on network representations concerns
the development of various methods and structures for cheaply
storing useful information about the network and making it readily
and conveniently accessible. This is particularly significant when
the network is large and geographically dispersed, and information
about its structure must be accessed from various local points in
it. As a notable example, the basic function of a communication
network, namely, message delivery, is performed by its routing
scheme, which requires maintaining certain topological knowledge.

Recently, a number of studies focused on a {\em localized} network
representation method based on assigning a (hopefully short) {\em
label} to each vertex, allowing one to infer information about any
two vertices {\em directly} from their labels, without using {\em
any} additional information sources. Such labeling schemes have been
developed for a variety of information types, including vertex
adjacency \cite{Breuer66,BF-67,KNR92}, distance
\cite{P99:lbl,KKP00,GPPR01,GKKPP01,GP01a,KM01,T01,CHKZ02,ABR03},
tree routing \cite{FG01,TZ01}, flow and connectivity \cite{KKKP02},
tree ancestry \cite{AKM01,AR02,KMS02}, nearest common ancestor in
trees \cite{AGKR01} and various other tree functions, such as
center, separation level, and Steiner weight of a given subset of
vertices \cite{Peleg00:lca}. See \cite{GP01b} for a survey.

By now, the basic properties of localized labeling schemes for {\em
static} (fixed topology) networks are reasonably well-understood. In
most realistic contexts, however, the typical setting is highly
dynamic, namely, the network topology undergoes repeated changes.
Therefore, for a representation scheme to be practically useful, it
should be capable of reflecting online the current up-to-date
picture in a dynamic setting. Moreover, the algorithm for generating
and revising the labels must be {\em distributed}, in contrast with
the sequential and centralized label assignment algorithms described
in the above cited papers.

The dynamic models investigated in this paper concern the {\em
leaf-dynamic} tree model in which at each step a leaf can be added
to or removed from the tree and the {\em leaf-increasing} tree model
in which the only topological event that may occur is that a leaf
joins the tree.
 We present a general method for constructing dynamic labeling
schemes which is based on extending  existing {\em static} tree
labeling schemes to the dynamic setting. This approach fits a number
of natural tree functions, such as routing , ancestry relation,
nearest common ancestor relation, distance and separation level.
Such an extension can be naively achieved by calculating the static
labeling from scratch after each topological change. Though this
method yields a good label size, it may incur
 a huge communication complexity.
Another naive approach would be that each time a leaf $u$ is added
as a child of an existing node $v$, the label given to $u$ is the
label of $v$ concatenated with $F(u,v)$. Such a scheme incurs very
little communication, however, the labels may be huge.

Before stating the results included in this paper, we list some
previous related works.% that deal with dynamic labeling schemes.
%%%%%%%%%%%%%%%%%%%%%%%%%%%%%%%%%%%%%%%%%%%%%%%%%%%%%%%%%%%%%%%%%
%%%%%%%%%%%%%%%%%%%%%%%%%%%%%%%%%%%%%%%%%%%%%%%%%%%%%%%%%%%%%%%%%
\paragraph*{\bf Related work:}
%%%%%%%%%%%%%%%%%%%%%%%%%%%%%%%%%%%%%%%%%%%%%%%%%%%%%%%%%%%%%%%%%
%%%%%%%%%%%%%%%%%%%%%%%%%%%%%%%%%%%%%%%%%%%%%%%%%%%%%%%%%%%%%%%%%
Static labeling schemes for routing on trees were investigated in
\cite{FG01}. For the {\em designer port} model, in which each node
can freely enumerate its incident ports, they show how to construct
a static routing scheme using labels of at most $O(\log n)$ bits on
$n$-node trees. In the {\em adversary port} model, in which the port
numbers are fixed by an adversary, they show how to construct a
static routing scheme using labels of at most $O(\frac{\log^2
n}{\log\log n})$ bits on $n$-node trees. They also show that the
label sizes of both schemes are asymptotically optimal.
Independently, a static routing scheme for trees using $(1+o(1))\log
n$ bit labels was introduced in \cite{TZ01} for the designer port
model.

A static labeling scheme for the id-based nearest common ancestor
(NCA) relation on trees was developed in \cite{Peleg00:lca} using
labels of $\Theta(\log^2 n)$ bits on $n$-node trees. A static
labeling scheme supporting the label-based NCA relation on trees
using labels of $\Theta(\log n)$ bits on $n$-node trees is presented
in \cite{AGKR01}.

In the {\em sequential} (non-distributed) model, dynamic data
structures for trees have been studied extensively (e.g.,
\cite{ST83,CH99,HLT01,AHT00}). For  comprehensive surveys on dynamic
graph algorithms see \cite{EGI99,FK00}.

Labeling schemes for the ancestry relation in the leaf-dynamic tree
model were investigated in \cite{CKM02}. They assume that once a
label is given to a node it remains unchanged. Therefore, the issue
of updates is not considered even for the non distributed setting.
For the above model, they present a labeling scheme that uses labels
of $O(m)$ bits, where $m$ is the number of nodes added to the tree
throughout the dynamic scenario. They also show that this bound is
asymptotically tight. Other labeling schemes are presented in the
above paper assuming that clues about the future topology of the
dynamic tree are given throughout the scenario.

The study of dynamic distributed labeling schemes was initiated by
\cite{KPR04}. Dynamic distributed distance labeling schemes on trees
were investigated in \cite{KPR04} and \cite{KP03}. In \cite{KPR04}
they present a dynamic labeling scheme for distances in the
leaf-dynamic tree model with message complexity $O(\sum_i\log^2
n_i)$, where $n_i$ is the size of the tree when the $i$'th
topological event takes place. The protocol maintains $O(\log^2 n)$
bit labels, when $n$ is the current tree size. This label size is
proved in \cite{GPPR01} to be asymptotically optimal even for the
static (unweighted) trees scenario.

In \cite{KP03} they develop two $\beta$-approximate distance
labeling schemes (in which
 given two labels, one can infer a
$\beta$-approximation to the distance between the corresponding
nodes). The first scheme applies to the {\em edge-dynamic} tree
model, in which the vertices of the tree are fixed but the (integer)
weights of the edges
 may change (as long as they remain positive). The second scheme
applies to the {\em edge-increasing} tree model, in which the only
topological event that may occur is that an edge increases its
weight by one. In scenarios where at most $m$ topological events
occur, the message complexities of the first and second schemes are
$O(m\Lambda\log^3 n)$ and $O(m\log^3 n+n\log^2 n\log m)$ ,
respectively, where $\Lambda$ is some density parameter of the tree.
The label size of both schemes is $O(\log^2 n+\log n\log W)$ where
$W$ denotes the largest edge weight in the tree.

The study of methods for extending static labeling schemes to the
dynamic setting was also initiated in \cite{KPR04}. There, they
assume the designer port model and consider two dynamic tree models,
namely, the leaf-increasing and the leaf-dynamic tree models. Their
approach fits a number of natural functions on trees, such as
distance, separation level, ancestry relation, id-based and
label-based NCA relation, routing (in both the adversary and the
designer port models) etc.. Their resulting dynamic schemes incur
 overheads (over the static scheme)
on the label size and on the communication complexity. Specifically,
given a static $F$-labeling scheme $\pi$ for trees , let $\LS({\pi},
n)$ be the maximum number of bits in a label given by $\pi$ to any
vertex in any $n$-node tree, and let $\MC({\pi}, n)$ be the maximum
number of messages sent by $\pi$ in order to assign the static
labels in any $n$-node tree. Assuming  $\MC({\pi}, n)$ is
polynomial\footnote{ \label{foot1} The actual requirement is that
the message complexity is bounded from above by some function $f$
which satisfies $f(a + b)\geq f(a) + f(b)$ and $f(\Theta(n)) =
\Theta(f(n))$. These two requirements are satisfied by most natural
relevant functions, such as $c \cdot n^\alpha \log^\beta n$, where
$c > 0$, $\alpha \geq 1$ and $\beta > 0$. For simplicity, we assume
$\MC(\cdot, n)$ itself satisfies these requirements.} in $n$, the
following dynamic schemes are derived. For the leaf-increasing tree
model, they construct a dynamic $F$-labeling scheme $\pi^{inc}$. The
maximum label given by $\pi^{inc}$ to any vertex in any $n$-node
tree is
 $O(\log n\cdot\LS({\pi}, n))$ and the number of messages sent by
$\pi^{inc}$ is $O(\log n\cdot\MC({\pi}, n))$. In the case where
$n_f$, the final number of nodes in the tree, is known in advance,
they construct a dynamic $F$-labeling scheme with label size
$O\left(\frac{\log n_f \log n}{\log\log n_f}\cdot\LS({\pi},
n)\right)$ and message complexity $O\left(\frac{\log n}{\log\log
n_f}\cdot\LS({\pi}, n_f)\right)$. For the leaf-dynamic tree model,
they construct two dynamic $F$-labeling schemes. Let $n_i$ be the
size of the tree when the $i$'th topological event takes place. The
first dynamic $F$-labeling scheme has label size $O(\log
n\cdot\LS({\pi}, n))$ and message complexity $O\left(\sum_i \log
n_i\cdot\frac{\MC({\pi},
  n_i)}{n_i}\right)+O(\sum_i \log^2 n_i)$ and the second  dynamic $F$-labeling scheme has
label size $O\left(\frac{\log^2 n}{\log\log n}\cdot\LS({\pi},
n)\right)$ and message complexity $O\left(\sum_i \frac{\log
n_i}{\log\log n_i}\cdot\frac{\MC({\pi},
  n_i)}{n_i}\right)+O(\sum_i \log^2 n_i)$.
In particular, for all the above mentioned functions, even if $n_f$
is known in advance, the best dynamic scheme of \cite{KPR04} incurs
$O(\sum_i \log^2 n_i)$  message complexity and overhead of $O(\log
n)$ over the  label size of the corresponding static scheme.

%%%%%%%%%%%%%%%%%%%%%%%%%%%%%%%%%%%%%%%%%%%%%%%%%%%%%%%%
\paragraph*{\bf Our contribution:}
%%%%%%%%%%%%%%%%%%%%%%%%%%%%%%%%%%%%%%%%%%%%%%%%%%%%%
Following \cite{KPR04}, we present a different method for
constructing dynamic labeling schemes in the leaf-increasing and
leaf-dynamic tree model. Our method is also based on extending
existing static labeling schemes to the dynamic setting. However,
our resulting dynamic schemes incur different trade-offs between the
overhead factors on the label sizes and the message communication.
In comparison to \cite{KPR04}, our trade-offs give better
performances for the label size, sometimes at the expense of
communication. Our approach fits the same class of tree functions as
described in \cite{KPR04}. The following results apply for both the
designer port model and the adversary port model.
 Given a static $F$-labeling scheme $\pi$ for trees, let
 $\LS({\pi}, n)$ and $\MC({\pi}, n)$ be as before.
Let  $k(x)$ be any reasonable\footnote{ \label{foot2} We require
that  $k(x)$, $\log_{k(x)} x$ and $\frac{k(x)}{\log k(x)}$ are
nondecreasing functions. Moreover we require that,
 $k(\Theta(x))=\Theta(k(x))$.
The above requirements are satisfied by most natural sublinear
functions such as $\alpha x^{\epsilon}\log^{\beta} x$,
$\alpha\log^{\beta}\log x$ etc..} sublinear function of $x$. For the
leaf-increasing tree model, we construct the dynamic $F$-labeling
scheme $\S^{k(x)}$. The maximum number of bits in a label given by
$\S^{k(x)}$ to any vertex in any $n$-node tree during the dynamic
scenario is $O(\log_{k(n)} n\cdot\LS({\pi}, n))$. The maximum number
of messages sent by $\S^{k(x)}$ in any dynamic scenario is
$O(k(n)\log_{k(n)}n\cdot\MC({\pi},n))$, where $n$ is the final
number of nodes in the tree.\\
In particular, by setting $k(n)=\log^{\epsilon} n$ for any
$\epsilon>0$, we obtain dynamic labeling schemes supporting all the
above mentioned functions, with message complexity
$O(n\frac{\log^{1+\epsilon} n}{\log\log n})$ and $O(\frac{\log
n}{\log\log n})$ multiplicative overhead over the corresponding
asymptotically optimal label size.

For the leaf-dynamic tree model, assuming $\LS(\pi,n)$ is
multiplicative\footnote{\label{foot3} We actually require that
$\LS(\cdot,n)$ satisfies
$\LS(\cdot,\Theta(n))=\Theta(\LS(\cdot,n))$. This requirement is
satisfied by most natural functions such as $c \cdot n^\alpha
\log^\beta n$, where $c > 0$, $\alpha \geq 0$ and $\beta > 0$.} we
construct the dynamic $F$-labeling scheme $\DL^{k(x)}$ with the
following complexities. The maximum number of bits in a label given
by $\DL^{k(x)}$ to any vertex in any $n$-node tree is $O(\log_{k(n)}
n\cdot\LS({\pi}, n))$ and the number of messages used by
$\DL^{k(x)}$ is $O\left(\sum_i k(n_i)(\log_{k(n_i)}
n_i)\frac{\MC({\pi},
 n_i)}{n_i}\right)+O(\sum_i \log^2 n_i)$,
where $n_i$ is the size of the tree when the $i$'th topological
event takes place. In particular, by setting $k(n)=n^{\epsilon}$ for
any $0<\epsilon<1$, we obtain dynamic labeling schemes with
asymptotically the same label size as the corresponding static
schemes and sublinear amortized message complexity. In particular,
we get dynamic labeling schemes with sublinear amortized message
complexity and asymptotically optimal label size for all the above
mentioned functions. Also, by setting $k(n)=\log^{\epsilon} n$ for
any $0<\epsilon<1$, we obtain dynamic labeling schemes supporting
all the above mentioned functions, with message complexity $O(\sum_i
\log^2 n_i)$ and $O(\frac{\log n}{\log\log n})$ multiplicative
overhead over the corresponding asymptotically optimal label size.
In contrast, note that for any of the above mentioned functions $F$,
the best dynamic $F$-labeling scheme of \cite{KPR04} (in the
leaf-dynamic model) has message complexity $O(\sum_i \log^2 n_i)$
and $O(\log n)$ multiplicative overhead over the corresponding
asymptotically optimal label size.
\paragraph*{\bf Paper outline:}
We start with preliminaries in Section 2. In Section 3 we present
the $\FS^k_p$ schemes which will be used in Section 4, where we
introduce the dynamic labeling schemes for the leaf-increasing and
the leaf-dynamic tree models. In Section 5 we discuss how to reduce
the external memory used for updating and maintaining the labels.
\section{Preliminaries}
Our communication network model is restricted to tree topologies.
The network is assumed to dynamically change via vertex additions
and deletions. It is assumed that the {\em root} of the tree, $r$,
is never deleted.
The following types of topological events are considered.\\
{\bf Add-leaf:} A new vertex $u$ is {\em added} as a child of an
existing vertex $v$.
Subsequently, $v$ is informed of this event.\\% and
{\bf Remove-leaf:} A leaf of the tree is {\em deleted}.
Subsequently, the leaf's parent is informed of this event.\\

We consider two types of dynamic models. Namely, the {\em
leaf-increasing} tree model in which the only topological event that
may occur is of type add-leaf, and the {\em leaf-dynamic} tree model
in which both types of topological events may occur.

Incoming and outgoing links at every node are identified by so
called {\em port-numbers}. When a new child is added to a node $v$,
the corresponding ports are assigned a unique port-number, in the
sense that no currently existing two ports of $v$ have the same
port-number. We consider two main variations, namely, the {\em
designer port} model and the {\em adversary port} model. The former
allows each node $v$ to freely enumerate its incident ports while
the latter assumes that the port numbers are fixed by an adversary.

Our method is applicable to any function $\func$ such that for every
two vertices $u$ and $v$ in the tree the following condition is
satisfied.
\begin{description}
\item{(C1)}
For every vertex $w$ on the path between $u$ and $v$, $\func(u, v)$
can be calculated in polynomial time from $\func(u, w)$ and
$\func(w, v)$.
\end{description}
In particular, our method can be applied to the ancestry relation,
the id-based and label-based  NCA relations and for the distance,
separation level and routing functions (both in the designer and the
adversary port models), thereby extending static labeling schemes
such as those of \cite{AGKR01,FG01,TZ01,Peleg00:lca,P99:lbl} to the
dynamic setting. We further assume, for simplicity of presentation,
that $\func$ is symmetric, i.e., $\func(u, v) = \func(v, u)$. A
slight change to the suggested protocols handles the more general
case, without affecting the asymptotic complexity results.

A {\em labeling scheme} $\pi=\langle \cM_{\pi},\cD_{\pi} \rangle$
for a function $\func$ on pairs of vertices of a tree is composed of
the following components:
\begin{enumerate}
\item
A {\em marker} algorithm $\cM_{\pi}$ that given a tree, assigns
labels to its vertices.
\item
A polynomial time {\em decoder} algorithm $\cD_{\pi}$ that given the
labels $L(u)$ and $L(v)$ of two vertices $u$ and $v$, outputs
$\func(u,v)$.
\end{enumerate}

In this paper we are interested in distributed networks where each
vertex in the tree is a processor. This does not affect the
definition of the decoder algorithm of the labeling scheme since it
is performed locally, but the marker algorithm changes into a {\em
distributed marker protocol}.

Let us first consider static networks, where no changes in the
topology of the network are allowed. For these networks we define
{\em static} labeling schemes, where the marker protocol $\cM$ is
initiated at the root of a tree network and assigns static labels to
all the vertices once and for all.

We use the following complexity measures to evaluate a static
labeling scheme $\pi=\langle\cM_{\pi},\cD_{\pi}\rangle$.
\begin{enumerate}
\item
{\em Label Size}, $\LS(\cM_{\pi}, n)$: the maximum number of bits in
a label assigned by $\cM_{\pi}$ to any vertex on any $n$-vertex
tree.
\item
{\em Message Complexity}, $\MC(\cM_{\pi}, n)$: the maximum number of
messages sent by $\cM_{\pi}$ during the labeling process on any
$n$-vertex tree. (Note that messages can only be sent between
neighboring vertices).
\end{enumerate}

We assume that the static labeling scheme assigns unique labels. For
any static labeling scheme, this additional requirement can be
ensured at an extra additive cost of at most $n$ to $\MC(n)$ and
$\log n$ to $\LS(n)$.

\begin{example}
\label{ex:1} The following is a possible static labeling scheme
$\SDFS$ for the ancestry relation on trees based on the notion of
interval schemes (\cite{SK85}, cf. \cite{Peleg00:book}). Given a
rooted tree, simply perform a depth-first search starting at the
root, assigning each vertex $v$ the interval $I(v)=[a, b]$ where $a$
is its DFS number and $b$ is the largest DFS number given to any of
its descendants. The corresponding decoder decides that $v$ is an
ancestor of $w$ iff their corresponding intervals, $I(v)$ and
$I(w)$, satisfy $I(v)\subseteq I(w)$. It is easy to verify that this
is a correct labeling scheme for the ancestry relation. Clearly,
$\MC(\SDFS, n) = O(n)$ and $\LS(\SDFS, n) = O(\log n)$.

Labeling schemes for routing are presented in \cite{FG01}. They
consider both the designer port model and the adversary port model.
The schemes of \cite{FG01} are designed as a sequential algorithm,
but examining the details reveals that these algorithms can be
easily transformed into  distributed protocols. In the designer port
model, we get a static labeling scheme for routing with label size
and message complexity similar to those of the $\SDFS$ static
labeling scheme. In the adversary port model we get a static
labeling scheme for routing with linear communication and
$O(\frac{\log^2 n}{\log\log n})$ label size. The label sizes of both
schemes are asymptotically optimal.
\end{example}

The dynamic labeling schemes involve a marker protocol $\cM$ which
is activated after every change in the network topology. The
protocol $\cM$ maintains the labels of all vertices in the
underlying graph so that the corresponding decoder algorithm will
work correctly. We assume that the topological changes occur
serially and are sufficiently spaced so that the protocol has enough
time to complete its operation in response to a given topological
change before the occurrence of the next change.

We distinguish between the label $\cM(v)$ given to each node $v$ to
deduce the required information in response to online queries, and
the additional external storage $Memory(v)$ at each node $v$, used
during updates and maintenance operations. For certain applications
(and particularly routing), the label $\cM(v)$ is often kept in the
router itself, whereas the additional storage $Memory(v)$ is kept on
some external storage device. Subsequently, the size of $\cM(v)$
seems to be a more critical consideration than the total amount of
storage needed for the information maintenance.

For the leaf-increasing tree model, we use the following complexity
measures to evaluate a dynamic labeling scheme
$\pi=\langle\cM_{\pi},\cD_{\pi}\rangle$.
\begin{enumerate}
\item
{\em Label Size}, $\LS(\cM_{\pi}, n)$: the maximum size of a label
assigned by the marker protocol $\cM_{\pi}$ to any vertex on any
$n$-vertex tree in any dynamic scenario.
\item
{\em Message Complexity}, $\MC(\cM_{\pi}, n)$: the maximum number of
messages sent by $\cM_{\pi}$ during the labeling process in any
scenario where $n$ is the final number of vertices in the tree.
\end{enumerate}

Finally, we consider the leaf-dynamic tree model, where both
additions and deletions of vertices are allowed. Instead of
measuring the message complexity in terms of the maximal number of
nodes in the scenario, for more explicit time references, we use the
notation $\vecn=(n_1, n_2, \ldots, n_f)$ where $n_i$ is the size of
the tree immediately after the $i$'th topological event takes place.
For simplicity, we assume $n_1 = 1$ unless stated otherwise. The
definition of  $\LS(\cM_{\pi}, n)$ remains as before, and the
definition of the
message complexity changes into the following.\\
{\em Message Complexity}, $\MC(\cM_{\pi}, \vecn)$: the maximum
number of messages sent by $\cM_{\pi}$ during the labeling process
in any scenario where $n_i$ is the size of the tree immediately
after the $i$'th topological event takes place.

\section{The finite semi-dynamic $F$-labeling schemes $\FS^k_p$}
In this section, we consider the leaf-increasing tree model and
assume that the initial tree contains a single vertex, namely, its
root. Given a static $F$-labeling scheme $\pi=\langle
\cM_{\pi},\cD_{\pi} \rangle$, we first fix some integer $k$ and
then, for each integer $p\geq 1$, we recursively define the dynamic
scheme $\FS^k_p$ which acts on growing trees and terminates at some
point. Each dynamic scheme $\FS^k_p$ is guaranteed to function as a
dynamic $F$-labeling scheme as long as it operates. It will follow
that Scheme $\FS^k_p$  terminates only when $n$, the number of nodes
in the current tree, is at least $k^p$. Moreover, the overheads
(over $\pi$) of Scheme $\FS^k_p$ are $O(p)$ on the label size and
$O(p\cdot k)$ on the message complexity. The $\FS^k_p$ schemes are
used in the next section as building blocks for our dynamic
$F$-labeling schemes. Let us first give an informal description of
the $\FS^k_p$ scheme and its analysis.

\subsection{Overview of Scheme $\FS^k_p$}
Scheme $\FS^k_p$ repeatedly invokes a {\em reset} operation on
different subtrees, in which the marker protocol of the static
labeling scheme is applied and the labels it produces are used to
construct the dynamic labels. It will follow that just before Scheme
$\FS^k_p$ terminates, a reset operation is invoked on the whole
current tree.

The $\FS^k_p$ schemes are defined recursively on $p$ as follows. In
Scheme $\FS^k_1$, whenever a new vertex joins the tree, a reset
operation is invoked on the whole tree, in which each vertex
receives the label given to it by the marker protocol of the static
labeling scheme. The decoder of Scheme $\FS^k_1$
 is simply the decoder algorithm of static
labeling scheme. Using a counter at the root, after $k$ such reset
operations, the scheme terminates.

Given Scheme $\FS^k_p$, we now define  Scheme $\FS^k_{p+1}$. We
start by running Scheme $\FS^k_p$ at the root, until it is supposed
to terminate. As mentioned before, just before Scheme $\FS^k_p$
terminates, a reset operation is invoked on $T_0$, the whole current
tree. This reset operation is referred to as a $(p+1)$-global reset
operation (it may also be referred to as an $l$-global reset
operation for other $l$'s). Before this  $(p+1)$-global reset
operation, the $\FS^k_{p+1}$ scheme is simply  the $\FS^k_{p}$
scheme (which is applied at the root). I.e., the label given to any
vertex $v$ by the $\FS^k_{p+1}$ scheme is the label given to $v$ by
the $\FS^k_{p}$ scheme, and the decoder of Scheme $\FS^k_{p+1}$ is
simply the decoder of Scheme $\FS^k_{p}$. During the above mentioned
$(p+1)$-global reset operation, each vertex $v\in T_0$ receives the
label $\cM_{\pi}(v)$ given to $v$ by the marker algorithm of the
static labeling scheme. Instead of terminating Scheme $\FS^k_p$, we
continue as follows. For every $v\in T_0$, let $T_v$ denote the
dynamic subtree rooted at $v$ that contains $v$ and $v$'s future
children as well as all their future descendants. After the above
mentioned $(p+1)$-global reset operation, each vertex $v\in T_0$
invokes Scheme  $\FS^k_p$ on $T_v$. If, at some point, one of these
$\FS^k_p$ schemes is supposed to terminate, instead of terminating
it, a reset operation (which is also referred to as a $(p+1)$-global
reset operation) is invoked on $T_0$, the whole current tree. Again,
after the above mentioned $(p+1)$-global reset operation, each
vertex $v\in T_0$ invokes Scheme  $\FS^k_p$ on $T_v$. As before, if,
at some point,  one of these $\FS^k_p$ schemes is supposed to
terminate, instead of terminating it, a ($(p+1)$-global) reset
operation is invoked on the whole current tree, and so forth. Using
a counter at the root, after $k$ such $(p+1)$-global reset
operations, the  $\FS^k_{p+1}$ scheme terminates.

After any of the above mentioned $(p+1)$-global reset operations,
the label given to a vertex $w\in T_v$ for some $v\in T_0$ contains
the following components. The label $\cM_{\pi}(v)$, the relation
$F(w,v)$ and the label given to $w$ by the $\FS^k_p$ scheme that is
applied on $T_v$. Given the labels $L(x)$ and $L(y)$ of two vertices
$x\in T_v$ and $y\in T_u$, where $v\neq u$, the decoder algorithm
finds $F(x,y)$ using {\bf 1)} the static decoder algorithm applied
on $\cM_{\pi}(v)$ and $\cM_{\pi}(u)$,  {\bf 2)} the relations
$F(x,v)$ and $F(y,v)$ and  {\bf 3)} the condition C1. If $x$ and $y$
are at the same subtree $T_v$, then the decoder finds  $F(x,y)$
using the decoder algorithm of the $\FS^k_p$ scheme applied on the
labels given to $x$ and $y$ by the  $\FS^k_p$ scheme (which was
invoked on $T_v$).

Using induction on $p$, it follows that Scheme $\FS^k_p$ may
terminate only when the number of nodes in the tree is at least
$k^p$. Also, using induction on $p$, it can be shown that the label
size of the dynamic scheme is at most $O(p)$ times the label size of
the static scheme $\pi$. The fact that the message complexity of
Scheme $\FS^k_p$ is $O(p\cdot k)$  times the message complexity of
$\pi$, intuitively follows from the following facts. 1) for every
$1\leq l\leq p$,  the different applications of Scheme $\FS^k_l$ act
on edge disjoint subtrees and 2) for every $1\leq l\leq p$, every
application of Scheme $\FS^k_l$ invokes an $l$-global reset
operations at most $k$ times.

 Scheme $\FS^k_p$ invokes Scheme $\FS^k_l$ for
different $l$'s  on different subtrees. These different applications
of Scheme $\FS^k_l$ induce a decomposition of the tree into subtrees
of different levels; an $l$-level subtree is a subtree on which
Scheme $\FS^k_l$ is invoked. In particular, the whole tree is a
$p$-level subtree and each vertex is contained in precisely one
$l$-level subtree, for each $1\leq l\leq p$. Moreover, subtrees of
the same level are edge-disjoint, however, subtrees of different
levels may overlap, in particular, for $1\leq l<p$, any $l$-level
subtree is (not necessarily strictly) contained in some $l+1$-level
subtree. Note that $l$-global reset operations can  be applied only
on $l$-level subtrees. The above mentioned decomposition of the tree
into subtrees is referred to as {\em the subtrees decomposition}. As
shown later, the subtrees decomposition  is quite different from the
tree decomposition (into bubbles) of \cite{KPR04}, on which their
dynamic schemes are based upon.

In order to add intuition, we now give a short informal description
of the $\FS^k_p$ scheme and the subtrees decomposition from a
non-recursive point of view. Initially, the root is considered as an
$l$-level subtree for every $1\leq l\leq p$. At any time, given the
current subtrees decomposition, Scheme $\FS^k_p$ operates as
follows. Whenever a leaf $v$ joins the tree as a child of vertex
$u$, for every $1\leq l\leq p$, the $l$-level subtree $T_l(u)$
becomes $T_l(u)\cup \{v\}$ and $T_l(v)$ is defined to be $T_l(u)$.
In addition, a (1-global) reset operation is invoked on $T_1(v)$.
This 1-global reset operation may result in a sequence of reset
operations as follows. If, after the last reset operation, the root
of $T_1(v)$ went through $k$ (1-global) reset operations then the
following happen.\\
 {\bf 1)} If, just before the reset operation, $T_2(v)$
strictly contained $T_1(v)$, then a (2-global) reset operation is invoked on $T_2(v)$, \\
{\bf 2)} $T_2(v)$ remains a 2-level subtree and $T_1(v)$ is no
longer considered as part of the subtrees
decomposition,\\
{\bf 3)}  each vertex $w\in T_2(v)$ becomes the root of a new
1-level
subtree, namely $T_w$. \\

In general, for every $1\leq l\leq p-1$, if after the last
$l$-global reset operation, the root of $T_l(v)$ went through $k$
($l$-global) reset operations then the following happen.\\
{\bf 1)} If, just before the reset operation, $T_{l+1}(v)$
strictly contained $T_{l}(v)$, then an ($(l+1)$-global) reset operation is invoked on $T_{l+1}(v)$, \\
{\bf 2)} $T_{l+1}(v)$ remains an $(l+1)$-level subtree but for every
$1\leq l'\leq l$, every subtree in the subtrees decomposition
containing an edge of $T_{l+1}(v)$ is removed from the subtrees decomposition,\\
{\bf 3)}  for every $1\leq l'\leq l$ and every vertex $w\in
T_{l+1}(v)$, the subtree $T_w$ is added to the subtrees
decomposition as a new
$l'$-level subtree. \\

If, after the last $p$-global reset operation, $T$ went through $k$
($p$-global) reset operation then Scheme $\FS^k_p$ terminates.

We are now ready to describe the $\FS^k_p$ scheme more formally.

\subsection{Scheme $\FS^k_p$}
We start with the following definition. A {\em finite semi-dynamic
$F$-labeling scheme}
 is a dynamic $F$-labeling scheme that is applied on
a dynamically growing tree $T$ and terminates at some point. I.e.,
the root can be in one of two states, namely, 0 or 1, where
initially, the root is in state 1 and when the root changes its
state to $0$, the scheme is considered to be terminated. The
requirement from a finite semi-dynamic $F$-labeling scheme
 is that until the root changes its state
to 0, the scheme operates as a dynamic $F$-labeling scheme. For a
finite semi-dynamic $F$-labeling scheme, $\cS$, we define its {\em
stopping time} $\TS(\cS)$ to be the minimum  number of nodes that
have joined the tree until the time $\cS$ terminates, taken over all
scenarios. Assuming $\TS(\cS)\geq n$, the complexities $\LS(\cS,n)$
and $\MC(\cS,n)$ are defined in the same manner as they are defined
for dynamic labeling schemes.

Let $\pi=\langle\cM_{\pi}\cD_{\pi}\rangle$ be a static $F$-labeling
scheme such that $\MC(\pi,n)$ is polynomial in $n$ (see footnote 1).
Fix some integer $k>1$. We now describe for each integer $p\geq 1$,
the finite semi-dynamic $F$-labeling scheme $\FS^k_p=\langle
\cM_p,\cD_p\rangle$.

Our dynamic schemes repeatedly engage the marker protocol of the
static labeling scheme, and use the labels it produces to construct
the dynamic labels. In doing so, the schemes occasionally apply to
the already labeled portion of the tree a {\em reset} operation
(defined below) invoked on some subtree $T'$.

\paragraph*{\bf Sub-protocol $\Reset(T')$}
\begin{itemize}
\item
The root of $T'$ initiates broadcast and convergcast operations (see
\cite{Peleg00:book}) in order to calculate $n(T')$, the number of
vertices in $T'$.
\item
The root of $T'$ invokes the static labeling scheme $\pi$ on $T'$.
\end{itemize}
We describe the finite semi-dynamic $F$-labeling schemes $\FS^k_p$
in a recursive manner. It will follow from our description that
Scheme $\FS^k_p$ terminates immediately after some Sub-protocol
$\Reset$ is invoked on the whole current tree, $T$. Throughout the
run of Scheme $\FS^k_p$, the root $r$ keeps a counter $\mu_p$. We
start by describing $\FS^k_1$.
\paragraph*{\bf Scheme $\FS^k_1$}
\begin{enumerate}
\item
If a new node joins as a child of the root $r$ then $r$ invokes
Sub-protocol $\Reset(T)$ on the current tree (which contains two
vertices).
\item
The root initializes its counter to $\mu_1=1$.
\item
If a new node joins the tree, it sends a signal to $r$ instructing
it to invoke Sub-protocol $\Reset(T)$ on the current tree $T$.
\item
 The root $r$ sets $\mu_1=\mu_1+1$. If  $\mu_1=k$ then $r$ changes it
state to 0 and the scheme terminates. Otherwise we proceed by going
back to the previous step.
\end{enumerate}
Clearly $\FS^k_1$ is a finite semi-dynamic $F$-labeling scheme.

Given the finite semi-dynamic $F$-labeling scheme
$\FS^{k}_{p-1}=\langle \cM_{p-1},\cD_{p-1}\rangle$, we now describe
the scheme $\FS^k_{p}=\langle \cM_{p},\cD_{p}\rangle$.

\paragraph*{\bf Scheme $\FS^k_{p}$}
\begin{enumerate}
\item
We first initiate $\FS^k_{p-1}$ at $r$.
 At some point during the scenario,
(after some application of Sub-protocol $\Reset(T)$), the root is
supposed to change its state to 0 in order to terminate Scheme
$\FS^k_{p-1}$. Instead of doing so, we proceed to Step 2.
\item
The root initializes its counter to $\mu_p=1$.

\item
Let $T_0$ be the tree at the last time  Sub-protocol $\Reset$ was
applied and let $\cM_{\pi}(u)$ be the static label given to $u\in
T_0$ in the second step of that sub-protocol.
\item
 If  $\mu_p=k$ then the root changes its
state to 0 and the scheme terminates. Otherwise we continue to the
next step.
\item
The root broadcasts a signal to all the vertices in $T_0$
instructing each vertex $u$ to invoke Scheme $\FS^k_{p-1}$ on $T_u$,
the future subtree rooted at $u$ which contains $u$ and $u$'s future
children as well as their future descendants. Let
$\FS_{p-1}(u)=\langle \cM^u_{p-1},\cD_{p-1}\rangle$ denote the
scheme $\FS^k_{p-1}$ which is invoked by $u$.
\item
For each vertex $w$, let $u$ be the vertex in $T_0$ such that $w\in
T_u$. The label given to $w$ by the marker $\cM_p$ is  defined as
$\cM_p(w)=\langle \cM_{\pi}(u),F(u,w),\cM^u_{p-1}(w)\rangle$.
\item
For a vertex $z$ and $i\in \{1, 2, 3\}$, let $L_i(z)$ denote the
$i$'th field of $L(z)$. Given two labels $L(x)$ and $L(y)$ of two
vertices $x$ and $y$, the decoder $\cD_{p}$ operates as follows.
\begin{itemize}
\item
If $L_1(x)=L_1(y)$ (which means that $x$ and $y$ belong to the same
subtree $T_u$ for some $u\in T_0$) then $\cD_{p}$ outputs
$\cD_{p-1}(L_3(x),L_3(y))$.% Otherwise:
\item
If $L_1(x)\neq L_1(y)$ then this means that $x\in T_u$ and $y\in
T_v$ where both $u$ and $v$ belong to $T_0$. Furthermore, $u$ is on
the path from $x$ to $v$ and $v$ is on the path from $x$ to $y$.
Therefore $F(x,u)=L_2(x)$, $F(u,v)=\cD_{\pi}(L_1(x),L_1(y))$ and
$F(v,y)=L_2(y)$. The decoder proceeds using Condition (C1) on
$\func$.
\end{itemize}
\item
If at some point during the scenario, some vertex $u\in T_0$ is
supposed to terminate $\FS_{p-1}(u)$ by changing its state to $0$,
then instead of doing so, it sends a signal to the root $r$ which in
turn invokes Sub-protocol $\Reset(T)$ and sets $\mu_p=\mu_p+1$. We
proceed by going back to Step 3.
\end{enumerate}

By induction it is easy to show that Scheme $\FS^k_p$ is indeed a
finite semi-dynamic $F$-labeling scheme. Let us first prove that the
stopping time of Scheme $\FS^k_p$ is at least $k^p$.
\begin{claim}
\label{stopping-time} $\TS(\FS^k_p)\geq k^p$.
\end{claim}
\begin{proof}
We prove the claim by induction on $p$. For $p=1$, it is clear from
the description of  Scheme $\FS^k_1$ that if this scheme terminates
then the number of nodes that have joined the tree is $k$. Assume by
induction that $\TS(\FS^k_p)\geq k^p$ and consider Scheme
$\FS^k_{p+1}$.

Recall that Scheme $\FS^k_{p+1}$ initially invokes (in Step 1)
Scheme $\FS^k_p$ until the latter is supposed to terminate. Then
(after some messages are sent), by Step 5 of Scheme $\FS^k_{p+1}$,
each vertex on the
 current tree invokes Scheme $\FS^k_p$ on its future
subtree until one of these schemes is supposed to terminate. If at
this point,
 Scheme $\FS^k_{p+1}$ does not terminate, then again (after some messages are sent),
 each vertex on the
 current tree invokes Scheme $\FS^k_p$ on its future
subtrees, on so forth. By Steps 2,4 and 8 of Scheme $\FS^k_{p+1}$,
when Scheme $\FS^k_{p+1}$ terminates, Step 5 has been applied $k-1$
times and Step 1 has been applied once. In each of these
applications of Scheme $\FS^k_p$ (which act on disjoint sets of
edges), by our induction hypothesis, at least $k^p$ vertices have
joined the corresponding subtree. Altogether, we obtain that at
least $k^{p+1}$ vertices have joined the tree. The claim follows.
\QED
\end{proof}

\begin{lemma}
\label{label-fin}
\begin{itemize}
\item
$\LS(\FS^k_p,n)= O(p\cdot\LS(\pi,n))$.
\item
$\MC(\FS^k_p,n)\leq 5pk\cdot \MC(\pi,n)$.
\end{itemize}
\end{lemma}
\begin{proof}
The existence of a static $F$-labeling scheme $\pi$ with labels of
at most $\LS(\pi,n)$ bits implies that for any two vertices $u$ and
$v$ in any $n$-node tree, $F(u,v)$ can be encoded using
$O(\LS(\pi,n))$ bits. This can be done by simply writing the labels
of the two vertices. The first part of the lemma follows by
induction.
We now turn to prove the second part of the lemma using induction on $p$.\\
Using the fact that  $\MC(\pi,a)\geq a$ for every $a\geq 1$, it
follows that for $p=1$, $\MC(\FS^k_1)\leq 5k\cdot \MC(\pi,n)$.
Assume by induction that $\MC(\FS^k_p)\leq 5pk\cdot \MC(\pi,n)$ and
consider Scheme $\FS^k_{p+1}$. We distinguish between two types of
messages sent by Scheme $\FS^k_{p+1}$
 during the dynamic scenario. The first type of messages consists
of the messages sent in the different applications of Scheme
$\FS^k_p$. The second type of messages consists of the broadcast
messages in Step 5 of Scheme $\FS^k_{p+1}$ and the messages resulted
from the applications of Step 8 of Scheme $\FS^k_{p+1}$ (which
correspond to sending a signal to the root and applying Sub-protocol
$\Reset$). Let us first bound from above the number of messages of
the first type. Recall that Scheme $\FS^k_{p+1}$ initially invokes
Scheme $\FS^k_p$ until the latter is supposed to terminate. Then
messages of the second type are sent
 and then each vertex on the
 current tree invokes Scheme $\FS^k_p$ on its future
subtree until one of these schemes is supposed to terminate. Again,
if at this point,
 Scheme $\FS^k_{p+1}$ does not terminate, then messages of the
 second type are sent and then
 each vertex on the
 current tree invokes Scheme $\FS^k_p$ on its future
subtrees, on so forth. Note that the different applications of
$\FS^k_p$ act on disjoint sets of edges and since we assume that
$\MC(\pi,(a+b))\geq \MC(\pi,a)+\MC(\pi,b)$ is satisfied for every
$a,b\geq 1$, we obtain (by our induction hypothesis) that the number
of messages of the first type is at most
 $5pk\cdot \MC(\pi,n)$. \\
By Steps 2,4 and 8 of Scheme $\FS^k_{p+1}$ we get that Step 3 of
Scheme $\FS^k_{p+1}$ can be applied at most $k$ times. Using the
fact that $\MC(\pi,a)\geq a$ for every $a\geq 1$, the total number
of messages of the second type  sent by Scheme $\FS^k_{p+1}$ is at
most $5k\MC(\pi,n)$. Altogether, we obtain that the number of
messages sent by Scheme $\FS^k_{p+1}$
 during the dynamic scenario is at most $5pk\cdot \MC(\pi,n)+
5k\cdot \MC(\pi,n)= 5(p+1)k\cdot \MC(\pi,n)$. The second part of the
lemma follows. \QED
\end{proof}

\subsection{The subtrees decomposition}
We refer to the a reset operations mentioned in either Step 1 or
Step 8 of the description of Scheme $\FS^k_p$ as a {\em $p$-global
reset operation}. Scheme $\FS^k_p$ invokes Scheme $\FS^k_l$ for
different $l$'s (where $1\leq l\leq p-1$) on different subtrees.
These different applications of Scheme $\FS^k_l$ induce a
decomposition of the tree into subtrees of different levels as
follows. At any time during the dynamic scenario, the whole tree is
considered as a $p$-level subtree. At any time before the first
$p$-global reset operation, the whole tree is also considered as a
$(p-1)$-level subtree. At any given time after the first $p$-global
reset operation, let $T_0$ denote the tree during the last
$p$-global reset operation. Between any two $p$-global reset
operations, the edges of $T_0$ are not considered as part of any
$l$-level subtree, where $l<p$, in other words, the edges of $T_0$
are only considered as part of the $p$-level subtree, which is the
whole tree. However, for each $v\in T_0$, the dynamic subtree $T_v$
is now considered as a $(p-1)$-level subtree. The decomposition into
subtrees induced by Scheme $\FS^k_p$ continues recursively using the
decomposition into subtrees induced by the $\FS^k_{p-1}$ schemes
which are applied on $T_v$ for every $v\in T_0$. We refer to the
resulting decomposition as {\em the subtrees decomposition}. The
following properties easily follow from the description of Scheme
$\FS^k_p$.\\
\subsubsection*{Subtrees decomposition properties}
\begin{enumerate}
\item
For any given $1\leq l\leq p$, the $l$-level subtrees are edge
disjoint.
\item
For every $1\leq l\leq p$, each vertex $v$ belongs to precisely one
$l$-level subtree; we denote this subtree by $T_l(v)$.
\item
Subtrees of different levels may overlap, in particular, any
$l$-level subtree, for $1\leq l<p$ is (not necessarily strictly)
contained in some $l+1$-level subtree.
\item
If $v$ is not the root of $T_l(v)$, then all $v$'s descendants also
belong to $T_l(v)$.
\item
Each reset operation may only be invoked on subtrees of the subtrees
decomposition.
\end{enumerate}

We note that, the dynamic schemes of \cite{KPR04} are based on the
{\em bubble tree decomposition} (see Subsection 4.1.1 of
\cite{KPR04}). As can be observed by the above properties, the
subtrees decomposition is quite different from the bubble tree
decomposition of \cite{KPR04}.

Since reset operations are carried on the subtrees of the subtrees
decomposition, each vertex $v$ must `know', for each $1\leq l\leq
p$, which of its incident edges belong to $T_l(v)$.
 The method by which each vertex $v$
implements the above is discussed in Section 5.

\section{The dynamic $F$-labeling schemes}
Let $\pi=\langle\cM_{\pi}\cD_{\pi}\rangle$ be a static $F$-labeling
scheme such that $\MC(\pi,n)$ is polynomial in $n$ (see footnote 1)
and let $k(x)$ be a sublinear function (see footnote 2). We first
construct the dynamic $F$-labeling scheme  $\S^{k(x)}$ for the
leaf-increasing tree model and then show how to transform it to our
 dynamic $F$-labeling scheme  $\DL^{k(x)}$ which is applicable
in the leaf-dynamic tree model.
\subsection{The dynamic $F$-labeling scheme $\S^{k(x)}$}
We  now describe our dynamic $F$-labeling scheme $\S^{k(x)}$ which
operates in the leaf-increasing tree model. Scheme $\S^{k(x)}$
invokes the $\FS^k_{p}$ schemes for different parameters $k$ and
$p$. Let us first describe the case in which the initial tree
contains a single vertex , i.e., its root. In this case, Scheme
$\S^{k(x)}$ operates as follows.
\paragraph*{\bf Scheme $\S^{k(x)}$}
\begin{enumerate}
\item
Invoke Scheme $\FS^{k(1)}_1$.
\item
Recall that while invoking Scheme $\FS^k_p$ , just before this
scheme is supposed to terminate, Sub-protocol $\Reset(T)$ is invoked
in which $n'$, the number of nodes in $T$, is calculated. For such
$n'$, let $p'$ be such that ${k(n')}^{p'}\leq 2\cdot
n'<{k(n')}^{p'+1}$. Let $p=p'+2$ and let $k=k(n')$. Instead of
terminating the above scheme, we proceed to the next step, i.e.,
Step 3 in Scheme $\S^{k(x)}$.
\item
The root of the whole tree invokes Scheme $\FS^k_p$ (with the
parameters $k$ and $p$ defined in the previous step) while ignoring
Step 1 of that scheme,
i.e., start directly in Step 2 of Scheme $\FS^k_p$.\\
At some point, Scheme $\FS^k_p$ is supposed to terminate. Instead of
terminating it, we proceed by going back to Step 2 of Scheme
$\S^{k(x)}$.
\end{enumerate}
\begin{theorem}
\label{sdl-theorem} $\S^{k(x)}$ is a dynamic $F$-labeling scheme for
the leaf-increasing tree model with the following complexities.
\begin{itemize}
\item
$\LS(\S^{k(x)},n)= O(\log_{k(n)} n\cdot\LS(\pi,n))$.
\item
$\MC(\S^{k(x)},n)= O(k(n)(\log_{k(n)} n) \MC(\pi,n))$.
\end{itemize}
\end{theorem}
\begin{proof}
At any given time $t$, there exist constants $k$ and $p$ such that
Scheme $\FS^k_p$ is applied by Scheme $\S^{k(x)}$. Let $n$ be the
current number of nodes in the tree and let $n'$ be the number of
nodes in the tree when the current Scheme $\FS^k_p$ was initiated.
We have $k^{p-2}(n')\leq 2\cdot n'<k^{p-1}(n')$ and therefore
$p-2\leq\log_{k(n')} 2n'$. Since $n'\leq n$ then by assumptions on
$k(x)$, we get that $p=O(\log_{k(n)}n)$ and the first part of the
theorem follows from the first part of Lemma \ref{label-fin}. We now
turn to prove the second part of the theorem.

For analysis purposes, we divide the scenario into sub-scenarios
according to the different applications of Step 3 in Scheme
$\S^{k(x)}$. We define these sub-scenarios as follows. Recall that
initially, Scheme $\S^{k(x)}$ invokes Scheme $\FS^{k(1)}_1$ until
the latter is supposed to terminate. We refer to the above mentioned
scenario as the 1'st scenario. For $i>1$, the $i$'th scenario
corresponds to the scenario between the $i-1$'st and the $i$'th
applications of Step 3 of Scheme $\S^{k(x)}$ (the $i$'th scenario
includes the $i-1$'st application of Step 3 and does not include the
$i$'th application of Step 3). Let $k_i$ and $p_i$ be the parameters
of the $\FS$ scheme corresponding to the $i$'th scenario and denote
this scheme by Scheme $\iFS$. Let $n_i$ be the number of nodes in
the tree at
the beginning of the $i$'th scenario.\\
{\bf Claim 1:}
For every $i>1$, $p_{i-1}\cdot k_{i-1}\leq (p_{i}-1)\cdot k_{i}$.\\
{\bf Proof:} Since Scheme $\FS^{k_{i-1}}_{p_{i-1}}$ was supposed to
terminate when Scheme $\iFS$ was initiated, Step 5 in Scheme
$\FS^{k_{i-1}}_{p_{i-1}}$ was applied $k_{i-1}-1$ times. Therefore,
by Claim \ref{stopping-time}, we obtain $(k_{i-1}-1)\cdot
k_{i-1}^{p_{i-1}-1}\leq n_{i}$  and therefore $k_{i-1}^{p_{i-1}}\leq
2\cdot n_{i}$, which implies $p_{i-1}\leq \log_{k_{i-1}} 2n_{i}$. We
therefore get that $p_{i-1}\cdot k_{i-1}\leq k_{i-1}\log_{k_{i-1}}
2n_i= \frac{k_{i-1}}{\log k_{i-1}}\log 2n_i$. By our assumption on
$k(x)$, we get that $\frac{k_{i-1}}{\log k_{i-1}}\log 2n_i\leq
\frac{k_i}{\log k_i}\log 2n_i=k_i\log_{k_i} 2n_i$.\\
By the choice of $k_i$ and $p_i$, We obtain that
$2n_i<k_i^{p_{i}-1}$ and therefore $\log_{k_i} 2n_i<p_{i}-1$.
Altogether, we obtain $p_{i-1}\cdot k_{i-1}\leq (p_i-1)\cdot k_i$,
as desired.\QED

\noindent{\bf Claim 2:} For any $i$, at any given time $t$ during
the $i$'th scenario, if the number of nodes in the tree at time $t$
is $n$, then the total number of messages sent by  $\S^{k(x)}$ until
time $t$ is at most
$5k_i p_i\MC(\pi,n)$.\\
{\bf Proof:} We prove the claim by induction on $i$. For $i=1$ we
have $k_1=k(1)$ and $p_1=1$ and the claim follows by the second part
of Lemma \ref{label-fin}. Assume that the claim is true for $i-1$
and consider a time $t$ in the $i$'th scenario such that the number
of nodes in the tree at time $t$ is $n$.

We distinguish between three types of messages sent until time $t$.
The first type of messages consists of the messages sent until the
$i$'th scenario was initiated. The second type of messages consists
of the messages sent in the different applications of Scheme
$\FS^{k_{i}}_{p_{i}-1}$ in Step 5 of Scheme $\iFS$. The third type
of messages consists of the broadcast messages resulting from Step 5
of Scheme $\iFS$ and the messages sent during the applications of
Step 8 of Scheme $\iFS$ (which correspond to sending a signal to the
root and applying Sub-protocol $\Reset$).

By our induction hypothesis and the previous claim,
 the number of messages of the first type is at most
$5k_{i-1} p_{i-1}\MC(\pi,n_{i})\leq 5k_{i}(p_{i}-1)\MC(\pi,n_{i})$.\\
By the second part of Lemma \ref{label-fin}, we get that if
$\FS^{k_{i}}_{p_{i}-1}$ is invoked on a growing tree whose current
number of nodes is $n'$, then the number of messages sent by
$\FS^{k_{i}}_{p_{i}-1}$ is at most $5k_{i}(p_{i}-1)\MC(\pi,n')$.
Using similar arguments as in the proof of Lemma  \ref{label-fin},
by our assumptions on $\MC(\pi,\cdot)$, we obtain that the total
number of messages of both the first type and the second type is at
most
 $5k_{i}(p_{i}-1)\MC(\pi,n)$.
Moreover, since Step 3 of Scheme $\iFS$ is applied at most $k_{i}$
times and since $\MC(\pi,a)\geq a$ for every $a\geq 1$, the total
number of messages of the third type is at most $5k_{i}\MC(\pi,n)$.
Altogether, we get that the number of messages sent by time $t$ is
at most $5k_{i}(p_{i}-1)\MC(\pi,n)+5k_{i}\MC(\pi,n)= 5k_{i}
p_{i}\MC(\pi,n)$ and the claim follows. \QED

Fix a time $t$ and let $n$ be the number of nodes in the tree at
time $t$. Let $i$ be such that time $t$ belongs to the $i$'th
scenario. By the choice of $k_i$ and $p_i$ and by our assumptions on
$k(x)$, we have $k_i=k(n_i)\leq k(n)$ and
$p_{i}=O(\log_{k(n_i)}n_i)=O(\log_{k(n)}n)$. The second part of the
theorem follows from Claim 2. \QED
\end{proof}

Let us now describe how to extend $\S^{k(x)}$ to the scenario in
which the initial tree $T_0$ does not necessarily  contains just the
root. In this case, Scheme $\S^{k(x)}$ operates as follows.
\paragraph*{\bf Scheme $\S^{k(x)}$ initiated on $T_0$}
\begin{enumerate}
\item
The root of $T_0$ invokes Sub-protocol $\Reset(T_0)$ in which
 the number of nodes $n_0$ in the initial tree is calculated.
 Let $p'$ be such that $k(n_0)^{p'}\leq 2\cdot n_0<k(n_0)^{p'+1}$.
Let $k=k(n_0)$ and let $p=p'+2$.
\item
The root invokes Scheme $\FS^k_p$ while ignoring Step 1 of that
scheme, i.e., start directly in Step 2 of Scheme $\FS^k_p$. At some
point, Scheme $\FS^k_p$ is supposed to terminate. Instead of
terminating it, we proceed by going to the next step, i.e., Step 3
of this scheme.
\item
Recall that while invoking Scheme $\FS^k_p$ , just before this
scheme is supposed to terminate, Sub-protocol $\Reset(T)$ is invoked
in which $n'$, the number of nodes in $T$, is calculated. For such
$n'$, let $p'$ be such that ${k(n')}^{p'}\leq 2\cdot
n'<{k(n')}^{p'+1}$. Let $p=p'+2$ and let $k=k(n')$. Instead of
terminating the above scheme, we proceed by going back to the
previous step, i.e., Step 2.
\end{enumerate}

The proof of the following theorem follows similar steps as the
proof of Theorem \ref{sdl-theorem}.

\begin{theorem}
\label{n_0-theorem} For any dynamic scenario in the leaf-increasing
tree model, where the initial number of nodes in the tree is $n_0$
and $n$ is the final number of nodes in the tree,
 $\S^{k(x)}$ is a dynamic $F$-labeling scheme, satisfying the following complexities.
\begin{itemize}
\item
$\LS(\S^{k(x)},n)= O(\log_{k(n)} n\cdot\LS(\pi,n))$.
\item
$\MC(\S^{k(x)},n)= O(k(n)(\log_{k(n)} n) \MC(\pi,n))$.
\end{itemize}
\end{theorem}

By examining the details in
\cite{AGKR01,FG01,Peleg00:lca,P99:lbl,KNR92} concerning the labeling
schemes supporting the above mentioned functions (i.e., the ancestry
relation, the label-based and the id-based NCA relations, the
separation level, the distance and the routing functions), it can be
easily shown that for each of the above mentioned labeling schemes
$\pi$, there exists a distributed protocol assigning the labels of
$\pi$ on static trees using a linear number of messages. Therefore,
by setting $k(n)=\log^{\epsilon} n$ for any $\epsilon>0$, we obtain
the following corollary.
\begin{corollary} In the leaf-increasing
tree model, there exist dynamic labeling schemes with message
complexity $O(n\frac{\log^{1+\epsilon} n}{\log\log n})$ for the
following functions.
\begin{itemize}
\item
For the routing function in the designer port model, the ancestry
relation and the label-based NCA relation: with label size
$O(\frac{\log^2 n}{\log\log n})$.
\item
For the routing function in the adversary port model: with label
size $O(\frac{\log^3 n}{\log^2\log n})$.
\item
For the distance function, the separation level and the id-based NCA
relation: with label size $O(\frac{\log^3 n}{\log\log n})$.
\end{itemize}
\end{corollary}

\subsection{The dynamic $F$-labeling scheme $\DL^{k(x)}$}

In the leaf-dynamic tree model, each vertex $u$ may store
information in $Memory(u)$ that is required for correct performances
of our dynamic schemes. One of the difficulties that may rise is
that when a leaf $u$ is deleted, we  lose the information stored in
$u$. In order to overcome this difficulty, we use the following
backup procedure. Throughout the dynamic scenario we maintain for
every child $u$ of a non-leaf node $v$,  a copy of $Memory(u)$
stored as backup in either $v$ or in a sibling of $u$. Thus, when
$u$ is deleted, $v$ retrieves the information in $Memory(u)$ by
communicating with the vertex holding the corresponding copy.
%
%5555555555555555555555555555555555555555555555555
% It turns out that the above mentioned backup procedure can be
% implemented without increasing the asymptotic label size and
% message complexity. The details regarding this implementation
%appear in ...............
%
%5555555555555555555555555555555555555555555555555555
 This is implemented as follows.

 Given a non-leaf node $v$, let $Ports(v)$ be the set port numbers
at $v$ leading to children of $v$ and let
 $u_i$ be the child of $v$ corresponding to the $i$'th smallest
port number in $Ports(v)$. Let $deg'(v)$ be the number of children
$v$ has. For a given child $u$ of $v$, let $index(u)$ be such that
$u=u_{index(u)}$ and let $next(u)$ be the child of $v$ satisfying
$index(next(u))=_{\mbox{mod } deg'(v)}index(u)+1$. Note that
$u=next(u)$ iff $deg'(v)=1$. Let $pre(u)$ be such that
$next(pre(u))=u$. Note that it only requires a local computation at
$v$ (and no extra memory storage) to detect for each $i$, which of
$v$'s port numbers leads to $u_i$.

The following invariants are maintained throughout
the dynamic scenario.\\
{\bf The copy invariants:}\\
{\bf 1)} For every child $u$ of a non-leaf node $v$,
 a copy of $Memory(u)$ is stored at either $v$ or $next(u)$.\\
{\bf 2)} Every vertex holds at most two such copies.\\

The copy invariants are maintained using the following steps applied
at each  node $v$.
\begin{enumerate}
\item
For every child $u$ of $v$, whenever the marker protocol of the
dynamic scheme updates $Memory(u)$ the  following happen. If
$deg'(v)>1$ then a copy of the new $Memory(u)$ is kept at $next(u)$
and the previous copy (if one exists) corresponding to a sibling of
$next(u)$ is erased from $next(u)$. If $deg'(v)=1$ then a copy of
the new $Memory(u)$ is kept at $v$ and the previous copy (if one
exists) corresponding to a child of $v$ is erased from $v$.
\item
If a child $u$ of $v$ is added to the tree then the  following
happen.  If $v$ was a leaf before $u$ was added then  a copy of the
new $Memory(u)$ is kept at $v$. Otherwise, if $u$ has other siblings
then a copy of the new $Memory(u)$ is kept at $next(u)$ and the
previous copy (if one exists) corresponding to a sibling of
$next(u)$ is erased from $next(u)$. In addition, a copy of
$Memory(pre(u))$ is kept at $u$.
\item
If a child $u$ of $v$ is removed from the tree then the following
happen.
\begin{enumerate}
\item
$v$ uses the copy of $Memory(u)$ (which is stored either at $v$ or
at $next(u)$) in order to perform the update tasks required by the
dynamic labeling scheme (this step is described in more detail in
the description of the corresponding dynamic scheme).
\item
If after the deletion, $deg'(v)=1$, then $v$ keeps a copy of
$Memory(w)$ for its only child $w$. In addition, the previous copy
(if one exists)  corresponding to a child of $v$ is erased from $v$.
\item
If after the deletion, $deg'(v)>1$, then $next(u)$ keeps a copy of
$Memory(pre(u))$ and the previous copy of $Memory(u)$ which was kept
at $next(u)$ is erased from $next(u)$.
\end{enumerate}
\end{enumerate}

The proof of the following lemma is straightforward.
\begin{lemma}
The copy invariants are maintained throughout the dynamic scenario.
\end{lemma}

Note that the asymptotic message complexity of the dynamic scheme is
not affected by the updates described above. Moreover, the second
copy invariant ensures that the asymptotic memory size of the scheme
remains the same.

Before turning to describe our main dynamic labeling scheme for the
leaf-dynamic model,
 let us first describe a version
of Scheme $\S^{k(x)}$, denoted $\sS^{k(x)}$, which operates in the
leaf-dynamic tree model and mimics the behavior of Scheme
$\S^{k(x)}$ on the dynamic scenario assuming deletions are never
made.
 Recall that in the leaf-increasing tree model, Scheme $\S^{k(x)}$
occasionally invokes sub-protocol $\Reset$ on different subtrees
$T'$ and that in the first step of this sub-protocol, the current
number of nodes in $T'$ is calculated. In the leaf-dynamic tree
model, Scheme $\sS^{k(x)}$ carries out the
same steps as $\S^{k(x)}$ except for the following two modifications.\\
{\bf 1)} Messages are not passed to deleted vertices.\\
{\bf 2)} Every time Scheme $\S^{k(x)}$ invokes Sub-protocol
$\Reset(T')$, instead of calculating the current number of nodes in
$T'$ in the first step of Sub-protocol $\Reset(T')$, Scheme
$\sS^{k(x)}$ calculates the number of nodes that have ever been in
$T'$, i.e., the existing
nodes in $T'$ together with the deleted ones.\\

The first modification is implemented trivially. Let us now describe
how to implement the second modification. Recall that by the
subtrees decomposition properties, each vertex $v$ belongs to
precisely one $l$-level subtree for each $1\leq l\leq p$. Moreover,
an $l$-global  reset operation is invoked only on $l$-level
subtrees. At any given time, let $T_l$ be some $l$-level subtree.
Let $n^-(T_l)$ denote the number of nodes that have been deleted
from $T_l$ and let $n(T_l)=|T_l|+n^-(T_l)$, i.e., the number of
nodes that have ever been in $T_l$. Throughout the dynamic scenario,
for every $1\leq l\leq p$, each vertex $v$ keeps a counter
$\omega_l(v)$ such that the following invariant is maintained at all
times for every $l$-level subtree $T_l$.\\
{\bf The $T_l$-invariant:} $\sum_{\{v\in T_l\}} \omega_l(v)=
n(T_l)$.\\

Assuming that for every $l$-level subtree, the $T_l$-invariant holds
at all times, we now show how to implement the second modification.
Instead of calculating the current number of nodes in $T_l$ in the
first step of Sub-protocol $\Reset(T_l)$, we calculate $n(T_l)$
using broadcast and upcast operations (see \cite{Peleg00:book}) on
$T_l$ by which $\sum_{\{v\in T_l\}} \omega_l(v)$ is calculated.

We now describe how Scheme $\sS^{k(x)}$ guarantees that the
$T_l$-invariant is maintained for every subtree $T_l$.

\begin{enumerate}
\item
If Scheme $\sS^{k(x)}$ is invoked on the initial tree $T_0$ then let
$p$ such that $\FS_p^k$ is initially invoked on $T_0$. For every
vertex $v\in T_0$, set $\omega_l(v)=1$ for every $1\leq l\leq p$.
\item
If $v$ is added as a leaf to the tree then $v$ sets $\omega_l(v)=1$
for every $1\leq l\leq p$.
\item
If $v$ participates in some Sub-protocol $\Reset$ which is invoked
by some Scheme $\FS^{k}_{l'}$ then for every $1\leq l< l'$, $v$ sets
$\omega_l(v)=1$.
\item
If a child $u$ of $v$ is deleted, then  $v$ extracts
$\{\omega_l(u)\mid 1\leq l\leq p\}$ using the copy of $Memory(u)$
(as mentioned before). % which is stored at either $v$ itself or at
%$next(u)$.
Subsequently, $v$ sets $\omega_l(v)=\omega_l(v)+\omega_l(u)$ for
every $1\leq l\leq p$ such that $T_l(v)=T_l(u)$.
\end{enumerate}

Using induction on the time, it is easy to verify that for every
subtree $T_l$, the  $T_l$-invariant is indeed maintained at all
times. Therefore, Scheme $\sS^{k(x)}$ can implement the
modifications to Scheme $\S^{k(x)}$ described above. Thus, using the
same steps as in the proof of Theorem \ref{n_0-theorem}, we obtain
the following lemma.
\begin{lemma}
\label{n_0+n^+} For any dynamic scenario in the leaf-dynamic tree
model, where the initial number of nodes in the tree is $n_0$ and
$n^+$ additions are made, Scheme
 $\sS^{k(x)}$ is
 a dynamic $F$-labeling scheme with the following complexities. Let
 $n'=n_0+n^+$.
\begin{itemize}
\item
$\LS(\sS^{k(x)},n)= O(\log_{k(n')} n'\cdot\LS(\pi,n'))$.
\item
$\MC(\sS^{k(x)},n')= O(k(n')(\log_{k(n')} n') \MC(\pi,n'))$.
\end{itemize}
\end{lemma}

 We now turn to describe Scheme
$\DL^{k(x)}$ which is designed to operate in the leaf-dynamic tree
model and improves the complexities of $\sS^{k(x)}$. Scheme
$\DL^{k(x)}$  uses a method similar to the one presented in
Subsection 3.4 in \cite{KPR04}. The general idea is to run, in
parallel to $\sS^{k(x)}$, a protocol for estimating the number of
topological changes in the tree. Every $\Theta(n)$ topological
changes we restart Protocol $\sS^{k(x)}$ again on the current
initial tree $T_0$.

 Denote by $\tau$ the number
of topological changes made to the tree during the execution in the
leaf-dynamic tree model. Fix $\delta = 9/8$. We use Protocol $\CW$
from \cite{KPR04} (which is an instance of the protocol of
\cite{AAPS96:jacm}) in which the root maintains an estimate
$\tilde\tau$ of $\tau$. This is done by applying the same mechanism
as in Protocol $\WW$ from \cite{KPR04} separately for the additions
of vertices and for the deletions. I.e, we run two protocols in
parallel. The first is designed to count the additions. In order to
do that, we ignore the deletions and perform the same steps as in
Protocol $\WW$. The second protocol is designed to count the
deletions. For this we ignore the additions, and carry the same
steps as in Protocol $\WW$, except for deletions rather than for
additions.
 Let $n_0$ be the number of vertices in the tree when
Protocol $\CW$ was initiated. Let $n_+$ and $n_-$ be the number of
additions and deletions respectively and let $\tilde n_+$ and
$\tilde n_-$ be the root's estimated number of additions and
deletions respectively.

As mentioned in Section 3.4.1 of \cite{KPR04}, as long as the root's
estimates satisfy $\tilde n_+\leq \frac {n_0}{9}$ and $\tilde
n_-\leq \frac {n_0}{9}$, it is guaranteed that $\tau=n_++n_-\leq
\frac {n_0}{2}$. Moreover, if $\tilde n_+>\frac {n_0}{9}$ or $\tilde
n^-> \frac {n_0}{9}$ then $\tau>\frac {n_0}{9}$. As mentioned in
\cite{KPR04}, $\MC(\CW, \vecn) = O(\sum_i \log^2 n_i)$.

Protocol $\DL^{k(x)}$ operates as follows.
\paragraph*{\bf Scheme $\DL^{k(x)}$}
\begin{enumerate}
\item \label{step1}
Let $T_0$ be the current tree. The root initiates a convergecast
process in order to calculate $n_0$, the initial number of nodes in
the tree.
\item
Protocols $\sS^{k(x)}$ and $\CW$ are started on $T_0$.
\item
When one of the estimates $\tilde n_+$ or $\tilde n_-$ exceeds
$\ninit/9$, return to Step \ref{step1}.
\end{enumerate}

\begin{theorem}
$\DL^{k(x)}$ is a dynamic $F$-labeling scheme for the leaf-dynamic
tree model, satisfying the following properties.
\begin{itemize}
\item
$\LS(\DL^{k(x)}, n)= O(\log_{k(n)} n\cdot\LS(\pi,n))$.
\item
$\MC(\DL^{k(x)}, \bar{n})= O\left(\sum_i {k(n_i)}(\log_{k(n_i)} n)
\frac{\MC(\pi,n_i)}{n_i}\right)+O(\sum_i\log^2 n_i)$.
\end{itemize}
\end{theorem}

\begin{proof}
Scheme $\DL^{k(x)}$ is restarted by returning to Step \ref{step1}
after $\tau$ topological changes, for $\tau = \Theta(n_0)$, where
$n_0$ is the last recorded tree size at Step \ref{step1}.
Consequently, the current tree size satisfies $n=\Theta(n_0)$ and
$n=\Theta(n_0+n^+)$ where $n^+$ is the number of additions made from
the last time Step 1 was invoked. Therefore, by the first part of
Lemma \ref{n_0+n^+} and by our assumptions on $k(n)$ and
$\LS(\pi,n)$, we obtain the first part of the theorem.

Let us now turn to prove the second part of the theorem. Let
$i_1,\ldots,i_m$ be the indices of the topological changes on which
Scheme $\DL^{k(x)}$ returns to Step \ref{step1}. Denote by $M_l$ the
number of messages resulting from the $l$'th time until the $l+1$'st
time Scheme $\sS^{k(x)}$ is applied in Step 2 of Scheme
$\DL^{k(x)}$. Clearly
$$
\MC(\DL^{k(x)}, \vecn) ~=~ \sum_{l=1}^m M_l~+\MC(\CW,\vecn)~.
$$
Since the number of changes relevant to $M_l$ is $\Theta (n_{i_l})$,
by our assumptions on $k(\cdot)$ and $\MC(\pi,\cdot)$, we obtain
$$
M_l ~\leq~    O\left( k(n_{i_l})(\log_{k(n_{i_l})} n)
\MC(\pi,n_{i_l})\right)~.
$$

Again, by our assumptions on $\MC(\pi,\cdot)$ and $k(\cdot)$ and we
actually have that
$$
M_l ~\leq~  O\left(\sum_{j =i_l}^{i_{l+1} - 1}
k(n_{j})(\log_{k(n_{j})} n) \frac{\MC(\pi,n_{j})}{n_j}\right)~.
$$

Therefore
$$
\sum_{l=1}^m M_l~\leq O\left(\sum_j k(n_{j})(\log_{k(n_{j})} n)
\frac{\MC(\pi,n_{j})}{n_j}\right)~.
$$
Since $\MC(\CW,\vecn)=O(\sum_i\log^2 n_i)$, the second part of the
theorem follows. \QED
\end{proof}

By setting $k(x)=n^{\epsilon}$ for any $0<\epsilon<1$, we obtain the
following corollary.
\begin{corollary}
\begin{itemize}
\item
In the leaf-dynamic tree model, for every static $F$-labeling scheme
$\pi$, there exists a dynamic $F$-labeling scheme with the same
asymptotic label size as $\pi$ and sublinear amortized message
complexity.
\item
In the leaf-dynamic tree model, there exist dynamic labeling schemes
for the ancestry relation, the id-based and label-based  NCA
relations and for the routing function (both in the designer and the
adversary port models) using asymptotically optimal label sizes and
sublinear amortized message complexity.
\end{itemize}
\end{corollary}
By setting $k(x)=\log^{\epsilon}n$ for any $0<\epsilon<1$, we obtain
the following corollary.
\begin{corollary}
\begin{itemize}
\item
In the leaf-dynamic tree model, for every static $F$-labeling scheme
$\pi$, there exists a dynamic $F$-labeling scheme with
$O\left(\sum_i \frac{\log^{1+\epsilon} n}{\log\log n}\cdot
\frac{\MC(\pi,n_i)}{n_i}\right)+O(\sum_i\log^2 n_i)$ message
complexity and multiplicative overhead of $O(\frac{\log n}{\log\log
n})$ over the label size of $\pi$.
\item
In the leaf-dynamic tree model, there exist dynamic labeling schemes
for all the above mentioned functions, with message complexity
$O(\sum_i\log^2 n_i)$ and multiplicative overhead of $O(\frac{\log
n}{\log\log n})$ over the  corresponding asymptotically optimal
label size.
\end{itemize}
\end{corollary}

\section {External memory complexity}
\subsection{Types of memory}
We distinguish between three types of memory bits used by a node
$v$. The first type consists of the bits in the label $\cM(v)$ given
to $v$ by the marker algorithm. The second type consists of the
memory bits used by the static algorithm $\pi$ in order to calculate
the static labels. The third type of bits , referred to as the {\em
external memory} bits, consists of the additional external storage
used during updates and maintenance operations by the dynamic
labeling scheme. As mentioned before,  for certain applications (and
particularly routing), the label $\cM(v)$ seems to be a more
critical consideration than the total amount of storage needed for
the information maintenance. In addition, the second type of memory
bits are used by the static algorithm $\pi$ only when it is invoked,
which is done infrequently. Moreover, we note that by
 examining the details in
\cite{AGKR01,FG01,Peleg00:lca,P99:lbl,KNR92} concerning the labeling
schemes supporting the ancestry relation, the label-based and the
id-based NCA relations, and the separation level, distance and
routing functions, it can be easily shown that for each of the above
mentioned labeling schemes $\pi$, there exists a distributed
protocol assigning the labels of $\pi$ on static trees using a
linear number of messages. Moreover, at any vertex, the number of
memory bits used by these static algorithms is asymptotically the
same as the label size.

In the following discussion, we therefore try to minimize the number
of external memory bits used by our dynamic schemes. Let us first
describe the need for these memory bits.

Consider either Scheme $\DL^{k(x)}$ or  Scheme $\S^{k(x)}$ for some
function $k(x)$. Recall that at any time during the dynamic
scenario, there exists parameters $k$ and $p$ such that the only
$\FS$ schemes that are currently invoked are of the form $\FS^k_l$
where $1\leq l\leq p$. Moreover, every vertex $v$ belongs to
precisely one $l$-level subtree, namely $T_l(v)$, for each $1\leq
l\leq p$. Therefore, each node $v$ holds at most $p$ counters of the
form $\mu_l$ and  at most $p$ counters of the form $\omega_l(v)$.
Since each such counter contains $O(\log n)$ bits we get that
holding these counters incurs $O(\log_{k(n)}n\cdot\log n)$ external
memory bits per node.

Since each node $v$ may participate at the same time in different
schemes $\FS^k_l$ for different $l$'s, $v$ must know, for each
$1\leq l\leq p$, which of its
edges correspond to  $T_l(v)$, its $l$-level subtree. %, the subtree containing
%$v$ on which Scheme $\FS^k_l$ is invoked.
Naively storing this information at $v$ may incur $\Omega(p\cdot n)$
bits of memory. Note that in Scheme $\FS^k_l$, each vertex $v$
either communicates with its parent in $T_l(v)$ or with all its
children in $T_l(v)$. Moreover, for each $l$, if $v$ is not the root
of $T_l(v)$ then its parent in
 $T_l(v)$ is its parent in $T$, namely, $parent(v)$.
Therefore, in order for $v$ to know, for each $l$, which of its
ports leads to its parent in $T_l(v)$, it is enough for it to know
which port leads to $parent(v)$ and for each $l$ to keep a bit,
indicating whether $v$ is the root of $T_l(v)$ or not. This costs
$O(p+\log n)$ memory bits. For each $1\leq l\leq p$, let $E_l(v)$ be
the port numbers (at $v$) corresponding to
 the edges connecting $v$ to its children in $T_l(v)$.
Note that $E_p$ is precisely the collection of all port numbers at
$v$ leading to $v$'s children in $T$.
 It is therefore enough to ensure that $v$ is able to
detect, for every $1\leq l<p$, which of its port numbers are in
$E_l(v)$.

We first consider our schemes in the designer port model, and then
discuss them in the adversary port model. In the designer port model
we show that the external memory bits used by a vertex do not exceed
the asymptotic label size of the corresponding dynamic scheme.
However, in the adversary port model, for a given static scheme
$\pi$, if the port numbers given by the adversary use many bits (in
comparison to the the label size of $\pi$), then the external memory
bits used by a vertex may exceed the asymptotic label size of the
corresponding dynamic scheme. Let us note that assuming the designer
port model, if the labels of the corresponding static labeling
scheme use the port numbers (e.g. the routing scheme of \cite{FG01}
for the designer port model) then we cannot re-enumerate the port
numbers to save external memory bits. In the context of this
section, we therefore consider such a scheme as operating in the
adversary port model. Let us note, however, that in the designer
port model, of all the above mentioned functions, the only static
scheme  which actually uses the port numbers to derive its labels,
is the routing scheme of \cite{FG01} for the designer port model.
Since this static routing labeling scheme uses port numbers with are
encoded using only $O(\log n)$ bits, the external memory complexity
of our corresponding dynamic routing labeling schemes is
asymptotically the same as the label size. (See Corollary
\ref{rout}).

 For every neighbor $u$ of a given
 vertex $v$, denote by $port(u)$ the current port number at $v$
 leading to $u$.
\subsection{External memory in the designer port model}
In the designer port model, in order to reduce the memory storage
used at each node, we exploit the fact that the tree $T_l(v)$ is a
subtree of $T_{l+1}(v)$. This is done in the following manner. For
every $1\leq l< p$, each node $v$ keeps a variable $a_l$ and
maintains an enumeration of its ports so that the following
invariant is maintained
at all times.\\
{\bf The designer-invariant:} For every $l=1,2,\cdots, p-1$,
$\{1,2\cdots ,a_l\}=E_l(v)$. In other words, $v$ maintains an
enumeration of its ports so that the port numbers from 1 to $a_l$
correspond
to the edges connecting $v$ to its children in $T_l(v)$.\\

In order to ensure that the designer-invariant is maintained at all
times, we follow the following steps.
\begin{enumerate}
\item
If Sub-protocol $\Reset$ is initiated on the whole tree $T$ and
$v\in T$ then let $p$ be the largest such that Scheme $\FS^k_p$ is
currently invoked on $T$. In this case, $v$ first sets the port
number leading to its parent to be $deg(v)$ and selects the port
numbers leading to its children in arbitrarily manner from 1 to
$deg(v)-1$. Then $v$ sets $a_l=0$ for every $1\leq l< p$.
\item
If $v$ is added as a leaf to the tree then $v$ sets
$port(parent(v))=1$ and then sets $a_l=0$ for every $1\leq l< p$.
\item
If $v$ participates in some Sub-protocol $\Reset$ which is invoked
by some Scheme $\FS^{k}_{p'}$ then for every $1\leq l< p'$, $v$ sets
$a_l=0$.
\item
If $u$ is added as a child of $v$ then $v$ increases all its port
numbers by 1. In addition $v$ sets $port(u)=1$, and for every $1\leq
l< p$, $a_l=a_l+1$.
\item
If a child $u$ of $v$ is deleted then for every child $w$ of $v$, if
$port(w)>port(u)$ then $v$ sets $port(w)=port(w)-1$. Moreover, for
every $a_l$ such that $a_l\geq port(u)$, $v$ sets $a_l=a_l-1$.
\end{enumerate}

The proof of the following lemma is straightforward.
\begin{lemma}
For every vertex $v$, the designer-invariant is maintained  at all
times.
\end{lemma}

Using the designer-invariant, the port numbers in $E_l(v)$ can
easily be identified by $v$ since they are precisely the port
numbers $1,2,\cdots, a_l$.

Since each vertex $v$ holds $O(p)$ counters and at most $p$
variables of the form $a_l$, and since each of these variables and
counters contains $O(\log n)$ bits, we obtain the following lemma.
\begin{lemma}
\label{mem-fin} In the designer port model, for any execution of
either Scheme $\S^{k(x)}$ in the leaf-increasing tree model or
Scheme $\DL^{k(x)}$ in the leaf-dynamic tree model, the maximal
number of external memory bits used by a vertex in any $n$-node tree
is $O(\log_{k(n)}n\cdot\log n)$.
\end{lemma}

\subsection{External memory in the adversary port model}

We first remark that in \cite{KPR04}, the designer port model is
assumed. Since
 port numbers are used in the labels given by the dynamic
schemes of \cite{KPR04}, applying their scheme in the adversary port
model may affect the label sizes of the schemes. Specifically, let
$\tau(n)$ be the maximum port number given by the adversary to any
node in any $n$-node tree, taken over all scenarios. Then the upper
bound on the label sizes of the general schemes proposed in
\cite{KPR04} changes from $O(d\log_d n\cdot\LS(\pi,n))$ to
$O(d\log_d n\cdot(\LS(\pi,n)+\log\tau(n)))$ (see Lemma 4.12 in
\cite{KPR04}). In contrast, applying our schemes in the adversary
port model may only  affect the external memory complexities. As
discussed before, it is enough to guarantee that each node $v$ knows
for each $l<p$ which of its port numbers is in $E_l(v)$. In the
designer port model, in order to achieve this, $v$ uses the fact
that $T_l(v)$ is a subtree of $T_{l+1}(v)$ to enumerate its port
numbers accordingly. In the adversary port model, however, $v$
cannot assign new port numbers, therefore a different strategy must
be used. The strategy we propose is that each node $v$ distributes
 the relevant information to its children in $T$ and
collects it back when needed. We assume that the ports at each node
$v$ are hardwired in such a way that $v$ is able to know for each
$i$, which of its port numbers is the $i$'th smallest port number.
Note that $E_p(v)$ is in fact the set of port numbers at $v$ leading
to $v$'s children. Let $u_i$ be the child of $v$ corresponding to
the $i$'th smallest port number in $E_p(v)$.

\subsubsection{Adversary port model in the leaf-increasing tree model}

In the leaf-increasing tree model, for each $i$, node $u_i$ keeps a
table, denoted $Table(u_i)$, containing $p-1$ fields. For every
$1\leq l<p$, let  $Table_l(u_i)$ denote the $l$'th field of
$Table(u_i)$. Each such field is either empty or contains a port
number in $E_l(v)$. In addition, for every $1\leq l<p$, $v$ keeps a
counter $c_l$ such that the following two invariants
are maintained throughout the execution.\\
{\bf The $l$'th counters invariant:}  $c_l=|E_l(v)|$.\\
{\bf The $l$'th tables invariant:}  $\bigcup_{i=1}^{c_l}
Table_l(u_i)=E_l(v)$.\\

In order to implement these invariants, the counters and tables are
initialized and updated according to the following.
\begin{enumerate}
\item
If Scheme $\S^{k(x)}$ is initiated on the initial tree $T_0$ and
$v\in T_0$ then let $p$ be such that Scheme $\FS^k_p$ is currently
invoked on $T_0$ by Step 3 of $\S^{k(x)}$. For every $1\leq l<p$
initialize $c_l=0$. In addition,
 for every child $u$ of $v$ and for every $1\leq l<p$, initialize
$Table_l(u)=\emptyset$.
\item
If $v$ is added as a leaf to the tree then for every $1\leq l<p$
initialize $c_l=0$.
\item
If $v$ participates in some Sub-protocol $\Reset$ which is invoked
by some Scheme $\FS^{k}_{p'}$ then for every $1\leq l< p'$, set
$c_l=0$ and for every $j\leq c_l$ set $Table_l(u_j)=\emptyset$.
\item
If $u$ is added as a child of $v$ then let $j$ be such that
$port(u)$ is the $j$'th smallest port number among $v$'s children,
i.e., $u=u_j$. For every $1\leq l<p$ do the following. If $j\leq
c_l$ then set $Table_l(u)=port(u)$. Otherwise, if $j>c_l$ then set
$Table_l(u_{c_l+1})=port(u)$. In either case, after the above
mentioned
 updates, $v$ sets $c_l=c_l+1$.
\end{enumerate}

\begin{lemma}
\label{invariants} For every $1\leq l<p$,
 the $l$'th counters and $l$'th tables invariants are maintained
throughout the execution.
\end{lemma}
\begin{proof}
We prove the lemma by induction on the time. Two initial cases are
considered. The first is when Scheme $\S^{k(x)}$ is initiated on a
tree $T_0$ and $v\in T_0$. Note that in this case, if $p$ is the
largest such that $\FS^k_p$ is currently invoked on $T_0$, then
 for every $1\leq
l<p$, $E_l(v)$ is empty. Therefore, after initializing $c_l=0$ and
$Table_l(u)=\emptyset$ for every $l<p$ and for every child $u$ of
$v$, both invariants are trivially satisfied.

In the other initial case,  $v$ is added to the tree as a leaf. In
this case,
 after initializing $c_l=0$ for every $l<p$, the invariants are
again trivially satisfied since $v$ has no children.

Assume by induction that both invariants are maintained until time
$t$. The only two events that may affect the parameters of the
invariants at time $t+1$ are when $v$ participates in some
Sub-protocol $\Reset$ which is invoked by some Scheme $\FS^k_{p'}$
or when a child $u$ of $v$ is added to the tree. In the first case,
for every $p'\leq l<p$, none of the parameters of the $l$'th
counters and $l$'th tables invariants is changed. Therefore, by our
induction hypothesis,  for every $p'\leq l< p$, both the  $l$'th
counters and $l$'th tables invariants are maintained. However, for
every $1\leq l<p'$, $E_l(v)$ becomes empty. Therefore, for every
$1\leq l<p'$, after setting $c_l=0$, the $l$'th counters invariant
is maintained. By the fact that for every $j\leq c_l$ we update
$Table_l(u_j)=\emptyset$, the $l$'th tables invariant is maintained
as well.

In the second case, after $u$ is added as a child of $v$, the
corresponding port number $port(u)$ belongs to $E_l(v)$ for every
$1\leq l<p$. Therefore, by our induction hypothesis and by the fact
that $c_l$ is raised by one, for every $1\leq l<p$,  the $l$'th
counters invariant is maintained. Fix $1\leq l<p$ and let $j$ be
such that $u=u_j$. In order to prove that the $l$'th tables
invariant is maintained as well, note that if $j\leq c_l$ then after
adding $u$ to the tree and before updating the tables and counters,
by our induction hypothesis, we have $\bigcup_{i=1}^{j-1}
Table_l(u_i)
  \cup  \bigcup_{i=j+1}^{c_{l}+1}Table_l(u_i)=E_l(v)\backslash\{port(u)\}$.
Therefore, after updating $Table_l(u_j)=Table_l(u)=port(u)$ and
setting $c_l=c_l+1$, we obtain $\bigcup_{i=1}^{c_l}
Table_l(u_i)=E_l(v)$ and therefore the $l$'th tables invariant is
maintained. If on the other hand $j>c_l$, then after adding $u$ to
the tree and before updating the tables and counters, by our
induction hypothesis, we have $\bigcup_{i=1}^{c_l}
Table_l(u_i)=E_l(v)\backslash\{port(u)\}$. Therefore,  after
updating $Table_l(u_{c_l+1})=port(u)$ and then setting $c_l=c_l+1$,
we obtain $\bigcup_{i=1}^{c_l} Table_l(u_i)=E_l(v)$. Therefore, the
$l$'th tables invariant is maintained also in this case. The lemma
follows by induction. \QED
\end{proof}

As mentioned before, if node $v$ wishes to communicate with its
children in $T_l(v)$, it must collect the port numbers in $E_l(v)$.
By the $l$'th tables invariant, this can be done by inspecting the
$l$'th field in the tables of its children $u_1,u_2,\cdots,
u_{c_l}$. Note that $v$ can identify $u_i$, as it only requires a
local computation at $v$ to find out which of its ports has the
$i$'th smallest port number. Since the number of nodes $v$ needs to
inspect is the same as the number of its children in $T_l(v)$, this
inspection does not affect the asymptotic message complexity of the
scheme. Moreover, the first two types of updates mentioned above can
be carried out during the run of Scheme $\S^{k(x)}$ without
requiring extra messages. In the third type of update, at most
$c_{p'}=|E_{p'}(v)|$ neighbors of $v$ are updated, therefore, the
number of messages incurred by this type of updates is at most the
number of messages incurred by the corresponding $\Reset$
sub-protocols. Therefore, the number of messages incurred by this
type of updates does not affect the asymptotic message complexity of
Scheme $\S^{k(x)}$. The fourth type of update incurs
$O(p)=O(\log_{k(n)}n)$ messages per topological change. Altogether,
the asymptotic message complexity of Scheme $\S^{k(x)}$ does not
change as a result of the updates and inspections mentioned above.

Since the number of bits in each table $Table(u_i)$ is at most
$O(p\cdot\log \tau(n))$, and since $v$ keeps $O(p)$ counters (of the
form $c_l$ and $\mu_l$) each containing $O(\log n)$ bits, we obtain
the following lemma.
\begin{lemma}
Assuming the adversary port model and the leaf-increasing tree
model, the maximal number of external memory bits used by a vertex
in Scheme $\S^{k(x)}$ is $O(\log_{k(n)}n\cdot(\log \tau(n)+\log
n))$.
\end{lemma}

\subsubsection{Adversary port model in the leaf-dynamic tree model}
In the leaf-dynamic tree model, in order to maintain the tables
invariants we do the following. As before, for each $1\leq l<p$, $v$
keeps the counter $c_l$ and each child $u$ of $v$ keeps the table
$Table(u)$. In addition, each child $u$ of $v$ keeps another table,
denoted $Pointers(u)$, which also contains $p-1$ fields. For every
$1\leq l<p$, the $l$'th field in $Pointers(u)$, $Pointers_l(u)$, is
either empty (if $port(u)\notin E_l(v)$) or contains the port number
$port(w)$ of the child $w$ of $v$ satisfying $Table_l(w)=port(u)$.
In other words, the following
invariants are maintained for every child $u$ of $v$ and every $1\leq l<p$.\\
{\bf The $l$'th pointers invariants:}\\
{\bf 1)} $Pointers_l(u)=\emptyset$ iff
$port(u)\notin E_l(v)$.\\
{\bf 2)} $Pointers_l(u)=port(w)$ iff
$Table_l(w)=port(u)$.\\

In the leaf-dynamic tree model, for every $1\leq l<p$, the $l$'th
counters, $l$'th pointers and $l$'th tables invariants are
maintained by initializing and updating the counters $c_l$ at $v$
and the tables $Table(u)$ and $Pointers(u)$ at each child $u$ of
$v$. This is done in the following manner. We note that the
initializations and updates of the counters $c_l$ and the tables
$Table(u)$ in the Steps 1-4 described below, are similar to the
initializations and updates done in the leaf-increasing case.

\begin{enumerate}
\item
If Scheme $\S^{k(x)}$ is initiated on the initial tree $T_0$ and
$v\in T_0$ then let $p$ be such that $\FS^k_p$ is currently invoked
on $T_0$ by Step 3 of Scheme $\S^{k(x)}$. For every $1\leq l<p$
initialize, $c_l=0$ and
 for every child $u$ of $v$ and every $1\leq l<p$, initialize
$Table_l(u)=Pointers_l(u)=\emptyset$.
\item
If $v$ is added as a leaf to the tree then for every $1\leq l<p$
initialize $c_l=0$.
\item
If $v$ participates in some Sub-protocol $\Reset$ which is invoked
by some Scheme $\FS^{p'}_{k'}$ then for every $1\leq l< p'$, set
$c_l=0$. Moreover, for every $1\leq l<p'$ and every $j\leq c_l$ set
$Table_l(u_j)=\emptyset$. In addition, for every child $u$ of $v$ in
$T_{p'}$ (on which Sub-protocol $\Reset$ is invoked) and for every
$1\leq l<p$, set $Pointers_l(u)=\emptyset$.
\item
If $u$ is added as a child of $v$ then let $j$ be such that
$port(u)$ is the $j$'th smallest port number among $v$'s children,
i.e., $u=u_j$. For every $1\leq l<p$ do the following. If $j\leq
c_l$ then set $Table_l(u)=Pointers_l(u)=port(u)$. Otherwise, if
$j>c_l$ then set $Table_l(u_{c_l+1})=port(u)$ and set
$Pointers_l(u)=port(u_{c_l+1})$. In any case, after the above
mentioned updates, $v$ sets $c_l=c_l+1$.
\item
If a child $u$ of $v$ is deleted from the tree then $v$ extracts
$Table(u)$ and $Pointers(u)$ using the backup copy of $Memory(u)$
which is kept at either $v$ itself or at $next(u)$ (see Subsection
4.2). Let $j$ be such that $u=u_j$. For every $1\leq l<p$ consider
two cases.
\begin{enumerate}
\item
If $j\leq c_l$ then by the table invariant, before $u$ is deleted,
$Table_l(u)\in E_l(v)$. Let
$x$ be such that $Table_l(u)=port(x)$ and consider the following two subcases.\\
{\bf Subcase 5.a.1:} $x=u$.\\
In this subcase, $v$ sets $c_l=c_l-1$.\\
{\bf Subcase 5.a.2:} $x\neq u$.\\
In this subcase, we do the following. If $Pointers_l(u)=\emptyset$
then set $Table_l(u_{c_l})=port(x)$ and set
$Pointers_l(x)=port(u_{c_l})$. If on the other hand
$Pointers_l(u)=port(w)$ for some child $w$ of $v$, then set
$Table_l(w)=port(x)$, $Pointers_l(x)=port(w)$ and $c_l=c_l-1$.
\item
If $j>c_l$ then if $Pointers_l(u)=port(w)$ for some child $w$ of
$v$, then let $y$ be such that $Table_l(u_{c_l})=port(y)$. First set
$Table_l(w)=port(y)$ and then set $Table_l(u_{c_l})=\emptyset$ and
$c_l=c_l-1$. In addition, if $y\neq u$ then set
$Pointers_l(y)=port(w)$.

\end{enumerate}
\end{enumerate}

\begin{lemma}
\label{invariants-2} For every $1\leq l<p$, the $l$'th  counters,
$l$'th pointers and $l$'th tables invariants are maintained
throughout the execution.
\end{lemma}

\begin{proof}
We prove the lemma by induction on the time. The analysis proving
that $l$'th  counters and $l$'th tables invariants are maintained
after Steps 1-4 in similar to the analysis in the proof of Lemma
\ref{invariants}.

Two initial cases are considered. The first is when Scheme
$\S^{k(x)}$ is initiated on the initial tree $T_0$ and $v\in T_0$.
Note that in this case, if $p$ is such that $\FS^p_k$ is currently
invoked on $T_0$, then
 for every $1\leq
l<p$, $E_l(v)$ is empty. Therefore, after initializing $c_l=0$ for
every $1\leq l<p$ and $Table_l(u)=Pointers_l(u)=\emptyset$ for every
child $u$ of $v$ and for every $1\leq l<p$, all three invariants are
trivially satisfied.

In the other initial case,  $v$ is added to the tree as a leaf. In
this case,
 after initializing $c_l=0$ for every $l<p$, the invariants are
again trivially satisfied since $v$ has no children.

Assume by induction that for every $1\leq l<p$, the $l$'th counters,
$l$'th pointers and $l$'th tables invariants are maintained until
time $t$. The only three events that may affect the parameters of
the invariants at time $t+1$ are when $v$ participates in some
Sub-protocol $\Reset$ which is invoked by some Scheme
$\FS^{p'}_{k'}$ or when a child $u$ of $v$ is either added to or
removed from the tree. In the first case, for every $p'\leq l< p$,
none of the parameters of the $l$'th counters, $l$'th pointers and
$l$'th tables  invariants is changed. Therefore, by our induction
hypothesis, the $l$'th counters, $l$'th pointers and $l$'th tables
invariants are maintained also in this case. However, for every
$1\leq l<p'$, $E_l(v)$ becomes empty. Therefore, after setting
$c_l=0$ for every $1\leq l<p'$, the $l$'th counters invariant is
maintained. Fix $1\leq l<p'$. By our induction hypothesis and by the
fact that for every $j\leq c_l$ we update $Table_l(u_j)=\emptyset$,
the $l$'th tables invariant is maintained as well. Let us now
consider the $l$'th pointers invariants. By our induction
hypothesis, the $l$'th pointers invariants are maintained at time
$t$. Therefore, for every child $u$ of $v$ where $u\notin T_{p'}$,
$Pointers_l(u)=\emptyset$ (since $port(u)\notin E_l(v)$). Since we
update $Pointers_l(u)=\emptyset$ for every child $u$ of $v$ which is
in $T_{p'}$, it follows that for every child $u$ of $v$,
$Pointers_l(u)=\emptyset$. Since $E_l(v)=\emptyset$, the $l$'th
pointers invariants are maintained as well.

If $u$ is added as a child of $v$, then for every $1\leq l<p$, the
corresponding port number $port(u)$ belongs  to $E_l(v)$. Fix $1\leq
l<p$. By our induction hypothesis and by the fact that $c_l$ is
raised by one, the $l$'th counters invariant is maintained. Let $j$
be such that $u=u_j$. In order to prove that the $l$'th tables
invariant is maintained, note that if $j\leq c_l$ then after adding
$u$ to the tree and before updating the tables and counters, by our
induction hypothesis, we have $\bigcup_{i=1}^{j-1} Table_l(u_i)
  \cup  \bigcup_{i=j+1}^{c_{l}+1}Table_l(u_i)=E_l(v)\backslash\{port(u)\}$.
Therefore, after updating $Table_l(u)=port(u)$ and setting
$c_l=c_l+1$, we obtain $\bigcup_{i=1}^{c_l} Table_l(u_i)=E_l(v)$ and
therefore the $l$'th tables invariant is maintained. If on the other
hand $j>c_l$, then after adding $u$ to the tree and before updating
the tables and counters, by our induction hypothesis, we have
$\bigcup_{i=1}^{c_l} Table_l(u_i)=E_l(v)\backslash\{port(u)\}$.
Therefore,  after updating $Table_l(u_{c_l+1})=port(u)$ and then
setting $c_l=c_l+1$, we obtain $\bigcup_{i=1}^{c_l}
Table_l(u_i)=E_l(v)$. Therefore, the $l$'th tables invariant is
maintained also in this case. Let us now prove that the $l$'th
pointers invariants are maintained for every child $x$ of $v$.
 By our induction hypothesis, the $l$'th pointers invariants
are maintained for every child $x\neq u,u_{c_l}$ of $v$. If $j\leq
c_l$ then the $l$'th pointers invariants are maintained also for $u$
and $u_{c_l}$ since $Table_l(u)=Pointers_l(u)=port(u)$. If on the
other hand, $j> c_l$ then the $l$'th pointers invariants are
maintained for $u$ and $u_{c_l}$ since $Table_l(u_{c_l+1})=port(u)$
and $Pointers_l(u)=port(u_{c_l+1})$. Altogether, the $l$'th pointers
invariants are maintained for every child $x$ of $v$.

If a child $u$ of $v$ is deleted from the tree, then fix $1\leq l<p$
and let $j$ be such that $u=u_j$. Let us first consider Case 5.a in
which $j\leq c_l$. In this case, by our induction hypothesis, before
$u$ is deleted, $Table_l(u)\in E_l(v)$. Let $x$ be such that
$Table_l(u)=port(x)$. If $x=u$ then after setting $c_l=c_l-1$, by
our induction hypothesis, the $l$'th counters and $l$'th tables
invariants are maintained. Moreover, by our induction hypothesis,
the $l$'th pointers invariants are maintained before $u$ is deleted,
and in particular, $Table_l(u)=port(u)$ implies that
$Pointers_l(u)=port(u)$. Therefore the $l$'th pointers invariants
are maintained as well after deleting $u$.

Consider now the case where $Table_l(u)=port(x)$ and $x\neq u$. If
$Pointers_l(u)=\emptyset$ then since the $l$'th pointers invariants
are maintained before $u$ is deleted, we have $u\notin E_l(v)$. Note
that since $j\leq c_l$, by the $l$'th tables invariant, before $u$
is deleted, we have $port(x)\in E_l(v)$. Therefore, after $u$ is
deleted and before the updates we have $\bigcup_{i=1}^{c_l-1}
Table_l(u_i)=E_l(v)\backslash\{port(x)\}$. It follows that after
setting $Table_l(u_{c_l})=port(x)$, the $l$'th counters and $l$'th
tables invariants are maintained. Moreover, by our induction
hypothesis, after setting $Pointers_l(x)=port(u_{c_l})$, the $l$'th
pointers invariants are maintained as well. If on the other hand
$Pointers_l(u)=port(w)$ for some child $w$ of $v$, then by the
$l$'th pointers invariants, before $u$ is deleted, $u\in E_l(v)$.
Therefore, by our induction hypothesis, after setting
$Table_l(w)=port(x)$ and $c_l=c_l-1$, the $l$'th counters and $l$'th
tables invariants are maintained. In addition, by our induction
hypothesis, after setting $Pointers_l(x)=port(w)$, the $l$'th
pointers invariants are maintained as well.

Let us now consider Case 5.b in which $j> c_l$. First note that if
$Pointers_l(u)=\emptyset$ then by the $l$'th pointers invariants,
before $u$ is deleted, $port(u)\notin E_l(v)$ and therefore none of
the parameters of the $l$'th invariants is changed. Assume,
therefore, that $Pointers_l(u)=port(w)$ for some child $w$ of $v$.
By our induction hypothesis, the $l$'th pointers invariants are
maintained before $u$ is deleted and therefore, before $u$ is
deleted, $port(u)\in E_l(v)$. By our induction hypothesis, the
$l$'th tables invariant is maintained before $u$ is deleted.
Therefore, after $u$ is deleted and before the updates are made, we
have $\bigcup_{i=1}^{c_l} Table_l(u_i)=E_l(v)\cup port(u)$. Let $y$
be such that $Table_l(u_{c_l})=port(y)$. After setting
$Table_l(w)=port(y)$, $Table_l(u_{c_l})=\emptyset$ and $c_l=c_l-1$,
we obtain $\bigcup_{i=1}^{c_l} Table_l(u_i)=E_l(v)$. Therefore, the
$l$'th counters and $l$'th tables invariants are maintained after
the updates are made. In addition, by our induction hypothesis, if
$u\neq y$ then after setting $Pointers_l(y)=port(w)$, the $l$'th
pointers invariants are maintained. If, on the other hand $u=y$ then
also $w=u_{c_l}$ and the $l$'th pointers invariants are maintained
also in this case. The lemma follows by induction. \QED
\end{proof}

As mentioned before, if node $v$ wishes to communicate with its
children in $T_l(v)$, it must collect the port numbers in $E_l(v)$.
By the $l$'th tables invariant, this can be done (similarly to the
leaf-increasing tree model case) by inspecting the $l$'th field in
the tables of its children $u_1,u_2,\cdots, u_{c_l}$. Since the
number of nodes $v$ needs to inspect is the same as the number of
its children in $T_l(v)$, then this inspection does not affect the
asymptotic message complexity of the scheme. Moreover, the first two
types of updates mentioned above can be carried out during the run
of Scheme $\S^{k(x)}$ without requiring extra messages. In the third
type of update, at most $c_{p'}=|E_{p'}(v)|$ vertices are updated,
therefore, the number of messages incurred by this type of updates
is at most the number of messages incurred by the corresponding
$\Reset$ sub-protocols. In particular, the number of messages
incurred by this type of updates does not affect the asymptotic
message complexity of Scheme $\S^{k(x)}$. The fourth and fifth types
of updates incur $O(p)=O(\log_{k(n)}n)$ messages per topological
change. Altogether, the asymptotic message complexity of Scheme
$\DL^{k(x)}$ does not change as a result of the updates and
inspections mentioned above.

Since for every child $u$ of $v$, the number of bits in each table
$Table(u)$ and each table $Pointers(u)$ is at most $O(p\cdot\log
\tau(n))$ and since $v$ keeps $O(p)$ counters%(of the form $c_l$ and
%$\mu_l$)
, each containing $O(\log n)$ bits, we obtain the following lemma.

\begin{lemma}
Assuming the adversary port model and the leaf-dynamic tree model,
 the maximal number of external memory bits
used by a vertex in Scheme $\DL^{k(x)}$ is $O(\log_{k(n)}n\cdot(\log
\tau(n)+\log n))$.
\end{lemma}

As mentioned before, in the designer port model, if the
corresponding static labeling scheme uses the port numbers, in the
context of saving external memory bits, we consider such a scheme as
operating in the adversary port model. However, in the designer port
model, the only static scheme of all the above mentioned functions
whose corresponding static scheme actually use the port numbers is
the routing scheme of \cite{FG01} for the designer port model. Since
this static scheme uses port numbers with are encoded using $O(\log
n)$ bits, we obtain the following corollary.
\begin{corollary}\label{rout}
Let $\pi$ be the static routing scheme of \cite{FG01} for the
designer port model. Then the maximal number of external memory bits
used by a vertex in either Scheme $\DL^{k(x)}$ or Scheme $\S^{k(x)}$
is $O(\log_{k(n)}n\cdot\log n)$ (which is asymptotically the same as
the label size of the corresponding dynamic scheme).
\end{corollary}

\def\thepage{}
{\small

\end{document}